\documentclass[apj]{emulateapj}
\submitted{Accepted for publication in the Astrophysical Journal October 6, 2008}
\shortauthors{RANGWALA ET AL.} \shorttitle{FABRY-PEROT SPECTROSCOPY OF GALACTIC BAR: KINEMATICS}
\begin{document}


\newcommand{\ca}         {\mbox{\rm \ion{Ca}{2}} $\lambda 8542$}
\newcommand{\kms}        {km s$^{-1}$}

\title{Fabry-Perot Absorption Line Spectroscopy of the Galactic Bar. I. Kinematics}
\author{Naseem Rangwala\altaffilmark{1}, T. B. Williams\altaffilmark{1,2} and K. Z. Stanek\altaffilmark{2,3}}
\altaffiltext{1}{Department of Physics and Astronomy, Rutgers University, 136 Frelinghuysen Road,
Piscataway, NJ 08854} \altaffiltext{2}{Visiting astronomer, Cerro Tololo Inter-American
Observatory, National Optical Astronomy Observatory, which are operated by the Association of
Universities for Research in Astronomy, under contract with the National Science Foundation.}
\altaffiltext{3}{Department of Astronomy, Ohio State University, Columbus, OH}

\begin{abstract}
We use Fabry-P\'{e}rot absorption line imaging spectroscopy to measure radial velocities using the
\ca \ line in 3360 stars towards three lines of sight in the Milky Way's bar: Baade's Window and
offset position at $(l,b) \simeq (\pm 5.0, -3.5)\degr$. This sample includes 2488 bar red clump
giants, 339 bar M/K-giants, and 318 disk main sequence stars. We measure the first four moments of
the stellar velocity distribution of the red clump giants, and find it to be symmetric and
flat-topped. We also measure the line-of-sight average velocity and dispersion of the red clump
giants as a function of distance in the bar. We detect stellar streams at the near and far side of
the bar with velocity difference $\gtrsim$ 30 \kms at $l = \pm 5$\degr, but we do not detect two
separate streams in Baade's Window. Our M-giants kinematics agree well with previous studies, but
have dispersions systematically lower than those of the red clump giants by $\sim 10$ \kms. For
the disk main sequence stars we measure a velocity dispersion of $\sim 45$ \kms \ for all three
lines-of-sight, placing a majority of them in the thin disk within 3.5 kpc of the Sun, associated
with the Sagittarius spiral arm. We measure the equivalent widths of the \ca \ line that can be
used to infer metallicities. We find indications of a metallicity gradient with Galactic
longitude, with greater metallicity in Baade's Window. We find the bulge to be metal-rich,
consistent with some previous studies.
\end{abstract}


\section{Introduction}
More than half of the disk galaxies in the universe have central bars \citep{freeman96}. N-body
simulations show that bars form naturally in disk galaxies, and play a crucial role in their
formation and evolution \citep{sellwood93}. There is now abundant evidence that our own Galaxy is
barred \citep[see review by][]{gerhard99}. One of the earliest indications came from the large
non-circular motions of gas in the inner galaxy measured by \citet{devac64} and subsequently by
\citet{liszt80}. But it was not until the 1990s that the presence of a bar in the Milky Way (MW)
was firmly established by the observations of a distinct peanut-shaped bulge in the COBE NIR light
distribution \citep{blitz91,cobe98}, a non-axisymmetric signature in the gas kinematics
\citep{binney91, weiner99}, OGLE photometry showing a magnitude offset of the bulge red clump
giants at positive and negative longitudes \citep{stanek97}, and a large optical depth of the
bulge to microlensing \citep{alcock00}. All these observations agree that the bar is inclined at
some orientation angle with respect to the Sun-Galactic center line with its the near end in the
first Galactic quadrant ($0\degr \leq l \leq 90\degr$).

Nonaxisymmetric models of the inner Galaxy have been produced in the last decade.
\citet{bissantz02} and \citet{dwek95} used COBE data to model the density distribution in the
Galactic bar. \citet{weiner99} measured the properties of the bar using gas kinematics.
\citet{rattenbury07b} used red clump giants to trace the mass distribution of the bar.
\citet{zhao96} and \citet{bissantz04} have taken initial steps towards building fully
self-consistent stellar dynamical models based on the available density and kinematic data.
Existing stellar kinematic observation are insufficient to fully constrain the models, and there
is significant variation in the estimates of the bar parameters such as orientation ($15\degr -
45\degr$)\footnote{Although most previous investigations find bar orientation angles between 20 --
30\degr, others \citep{weiner99, benjamin05,cl08} have shown that the orientation angle could be
as high as 35\degr -- 45\degr.} , co-rotation radius (3.5 -- 5 kpc) and pattern speed ($35 - 60$
\kms \ kpc$^{-1}$). There is not yet a unified quantitative dynamical model of the Galactic bar
that explains all the observations in a consistent manner and the properties of the Galactic bar
are still poorly determined.
\begin{figure}[t]
\includegraphics[scale=0.37]{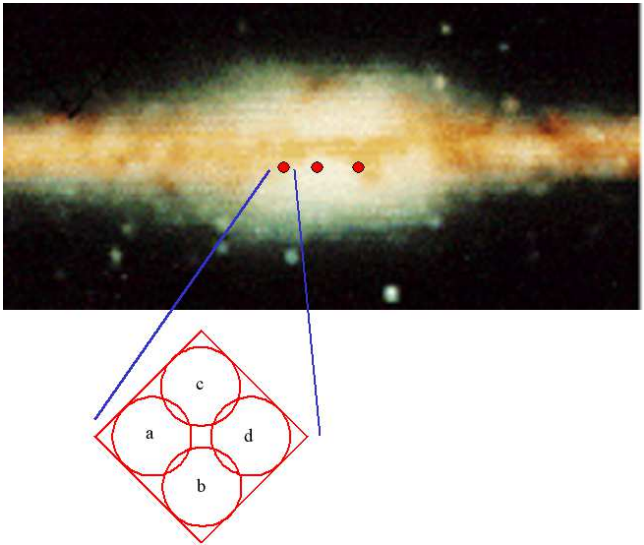}\\
\caption{The position of our fields against the COBE NIR background. From left to right are the
MM7B, BW and MM5B fields. MM7B and MM5B each have four overlapping sub-fields as shown in the
box.} \label{cobe}
\end{figure}

The signature of the Galactic bar should also be seen in the kinematic distortions of the stellar
velocity field, since the orbits are no longer circular, but rather are elongated streams along
the bar. The stellar kinematic evidence of bulge triaxiality is still ambiguous. There has been
significant work done in the past on the line-of-sight velocities of M and K giants \citep{rich90,
sharples, terndrup, minniti92, habing} and proper motions \citep{span, sumi04, rattenbury07a,
clarkson08} in the bulge. But most of these radial velocity observations were limited to the low
extinction Baade's Window ($l \sim 0$\degr). \citet{hafner} show that measurements in Baade's
Window (BW) alone are insensitive to the effects of triaxiality, and a stronger kinematic
signature of the bar will be apparent away from $l = 0$\degr. The ideal observation would be to
measure stellar velocity distributions with high S/N for several lines-of-sight (LOS) in the bar.
A sample size of $\sim 1000$ stars per LOS would enable the measurements of the higher moments of
the velocity distribution and of the average radial velocity shift between the stellar streams in
the bar \citep{mao02,deguchi01}. In addition, combining accurate proper motions with these LOS
velocities would provide even stronger constraints on the current dynamical models of the Galactic
bar.

Further, the ratio of organized stellar streaming motion around the bar to random motion
constrains the bar's angular rotation speed (also called pattern speed) \citep{sanders82}. Since
bars can interact strongly with dark matter halos, transferring angular momentum into the halo and
slowing the bar rotation dramatically \citep{victor00}, constraining the current pattern speed of
the MW bar will provide a bound on the disk to halo mass ratio for the inner Galaxy.

The most recent bulge radial velocity survey by \citet{howard08} \citep[also see][]{rich07}
measured radial velocities at many positions along the major and minor axes of the bar, obtaining
the rotation and velocity dispersion curves. These observations have addressed the crucial problem
of measuring kinematics of the bar away from BW. However, their sample size along each LOS is
limited to about 100 stars, insufficient to measure the higher moments of the individual velocity
distributions or the kinematics of the stellar streams. Our goal in the current investigation is
to obtain a large radial velocity sample along several LOS, in order to measure these important
quantities.

The basic procedure in Fabry-P\'{e}rot (FP) imaging spectroscopy is to obtain a series of
narrow-band images, tuning the interferometer over a range of wavelengths covering a spectral
feature of interest. The data cube thus produced is analyzed to extract a short portion of the
spectrum of each object in the field of view. This spectrum is fit to determine the strength,
width and central wavelength of the spectral feature, and the level of the adjacent continuum. If
there are many objects of interest within the field of view, the technique can be extremely
efficient. FP spectroscopy has traditionally been used to measure emission features in extended
objects, and has been highly successful for studying the velocity fields in disk galaxies
\citep[e.g.][]{palunas, zanmar08} star forming regions \citep{hartigan00}, planetary nebula
\citep{sluis06}, and supernova remnants \citep{ghavamian03}. It is less common to employ FP
techniques to measure absorption spectra, but they have been used to determine the kinematics of
stars in a large number of globular clusters \citep{gebhardt94}, and the pattern speed of the bar
in the early type galaxy NGC 7079 \citep{victor04}.
\begin{figure}[t]
\centering
 \includegraphics[scale=.67]{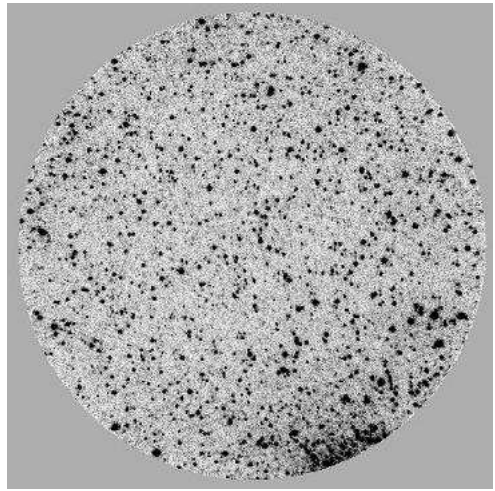}
 \caption{A Fabry-P\'{e}rot image of the Baade's Window stellar field.
 The FOV is $4^{\prime}$ and the central wavelength is 8543\,\AA. The increasing
  stellar density towards the southwest edge of the field is caused by the bulge globular cluster NGC 6522.
 \label{stellarfield}}
\end{figure}

In the present paper we use FP absorption line spectroscopy to obtain radial velocities using the
\ca \ line for 2488 Red Clump Giants (RCGs), 339 red giants (M/K type) and 318 disk main sequence
stars, towards three LOS in the MW bar: BW and two symmetric offset positions at $(l,b) \simeq
(\pm 5.0, -3.5)\degr$ (see Figure \ref{cobe}). This sample is an order of magnitude larger than
any previous sample for a given LOS. We also measure equivalent widths of the \ca \ absorption
lines that can be used to infer metallicities \citep[e.g.][]{az88} for the late type stars. This
will be the subject of a subsequent paper (Paper II) and is discussed only briefly here. In Paper
III we will use our measurements for new, more detailed dynamical models of the MW bar.
\begin{table*}
 \begin{center}
 \caption{Sub-fields observed in the MW bar. Right ascension ($\alpha$) and declination ($\delta$) are
 from epoch J2000 and $N_{stars}$ indicates the total number of stars (including foreground) obtained from each sub-field.}
 \begin{tabular}{llccccc} \tableline \tableline
 Field & Date Obs. & $\alpha$ & $\delta$ & $l(\degr)$ & $b(\degr) $
  & $N_{stars}$ \\
 \tableline
MM5B-a & 1996 & 17:47:24.05 & -35:00:21.07 & -5.00 & -3.47 & 472\\
MM5B-b & 1997 & 17:47:40.03 & -35:00:34.0 & -4.97 & -3.52 & 378\\
MM5B-c & 1999  & 17:47:20.71 & -35:56:52.84 & -4.95 & -3.43 & 608\\
MM5B-d & 1997/99 & 17:47:39.80 & -35:56:47.85 & -4.92 & -3.49 & 525\\
BWC-a & 1999 &  18:03:39.70 & -29:59:51.2 & 1.06 & -3.93 & 718\\
BW1-a & 1999  & 18:02:29.41 & -29:49:27.4 & 1.09 & -3.62 & 359\\
MM7B-a & 1996  & 18:11:29.61 & -25:54:06.6 & 5.50 & -3.46 & 353\\
MM7B-b & 1997  & 18:11:44.23 & -25:54:26.0 & 5.52 & -3.51 & 322\\
MM7B-c & 1997/99 & 18:11:30.32 & -25:50:38.5 & 5.55 & -3.44 & 288\\
MM7B-d & 1999 & 18:11:44.72 & -25:50:33.4 & 5.58 & -3.48 & 396\\
\tableline
 \end{tabular}
 \label{obs}
 \end{center}
\end{table*}

\section{Observations}
Observations were made between 1996 and 1999 using the Rutgers Fabry-P\'{e}rot (RFP) system at the
f/13.5 Cassegrain focus of the Cerro Tololo Inter-American Observatory (CTIO) 1.5m telescope. The
field of view of the RFP was circular with a diameter of $4^{\prime}$. The detector was a Tek 1024
x 1024 pixel CCD, giving an image scale of $0.36\arcsec$ per pixel. Typical seeing was in the
range of $1\arcsec - 1.5 \arcsec$. Each narrow-band exposure was of 900~s duration. A typical
image is shown in Figure \ref{stellarfield}. We observed three LOS: the MM5B field at l = -5\degr,
the MM7B field at l = +5\degr, and BW. There are four overlapping subfields in each of the MM5B
and MM7B fields, as shown in Figure \ref{cobe}, and two sub-fields in BW. Table \ref{obs} lists
the location of each sub-field and the total number of stars obtained per field. The choice and
nomenclature of the fields was motivated by the OGLE survey \citep{udalski}, which provides us
with excellent I and V band stellar photometry. In total we obtained $\sim 4400$ stellar spectra.

We chose to observe the \ca \ spectral line. The calcium triplet (8498.02~\AA, 8542.09~\AA\, and
8662.14~\AA) is one of the strongest near-infrared spectral features in late-type stars like the
RCGs. The triplet arises from the transition between the two levels of $3^2\textrm{D}$ and the two
levels of $4^2\textrm{P}$ in \ion{Ca}{2}. The temperature of the bulk of the ISM is not high
enough to excite \ion{Ca}{2} to the $3^2\textrm{D}$ state, and therefore no contamination from the
foreground ISM is expected. The 8542 \AA\ line is the strongest of the triplet and falls in a
wavelength region where the night sky is relatively free of strong terrestrial emission lines.

We used a medium-resolution etalon with a spectral response function that is well fit by a Voigt
function with a FWHM of 4~\AA, equivalent to approximately 140 \kms\ at 8500~\AA. We typically
scanned a range of $8530 - 8555$ \AA\ with wavelength steps of 1~\AA. The FP images are not
monochromatic, because the rays from off-axis points pass through the collimated beam at an angle,
making the passband shift to the blue towards the edge of the field. The scanning range was
sufficient to compensate for the 4\AA\ center-to-edge gradient by including extra images at the
red end of the scan. The resulting spectra allowed measurement of the central wavelength of the
line with high precision (5 -- 10 \kms).

This project was plagued by bad weather and technical difficulties with the instrument and
telescope, requiring a significantly longer period of time to acquire the data than expected.
During two of the nights in the 1997 run the instrument shutter malfunctioned, affecting the data
for sub-fields MM5B-b and MM7B-b, which upon reduction showed large deviations in kinematics
compared to the neighboring fields that overlap with them. For this reason we excluded these two
fields from our analysis. Because of time constraints only 18 images were obtained for the BW1-a
field with heavy obscuration from clouds for half of them, and hence we also excluded this field.
These exclusions brought the total sample size down to 3360 stars from seven sub-fields, of which
3144 are used in our kinematic analysis.

We calibrated the wavelength response of the system with several lines around 8542\,\AA\ from a
neon spectral lamp. The complete calibration sequence was done once per observing run, and was
stable over the duration of the run (and indeed from year to year). The precision of the
wavelength calibration was typically better than $0.05$~\AA, corresponding to 1.8 \kms at
8500~\AA. The zero-point of the wavelength scale was affected by changes in the environmental
conditions such as temperature and atmospheric pressure, and was monitored by taking a single neon
calibration image once each hour during the observations. The wavelength zero-point for each
observation image was determined by interpolation from these hourly calibrations. On images that
have night sky lines within their spectral range, we can also use these sky lines to check the
zero point. There were two weak OH lines (at 8548.71~\AA\ \& 8538.67~\AA; \citet{osterbrock}) that
fall within our scanning range. Although these lines had very low S/N, the zero-point we measured
from them for a few frames was consistent with the interpolated calibration ring measurement.
\section{Data Reduction}
\begin{figure*}
 \includegraphics[scale=0.9]{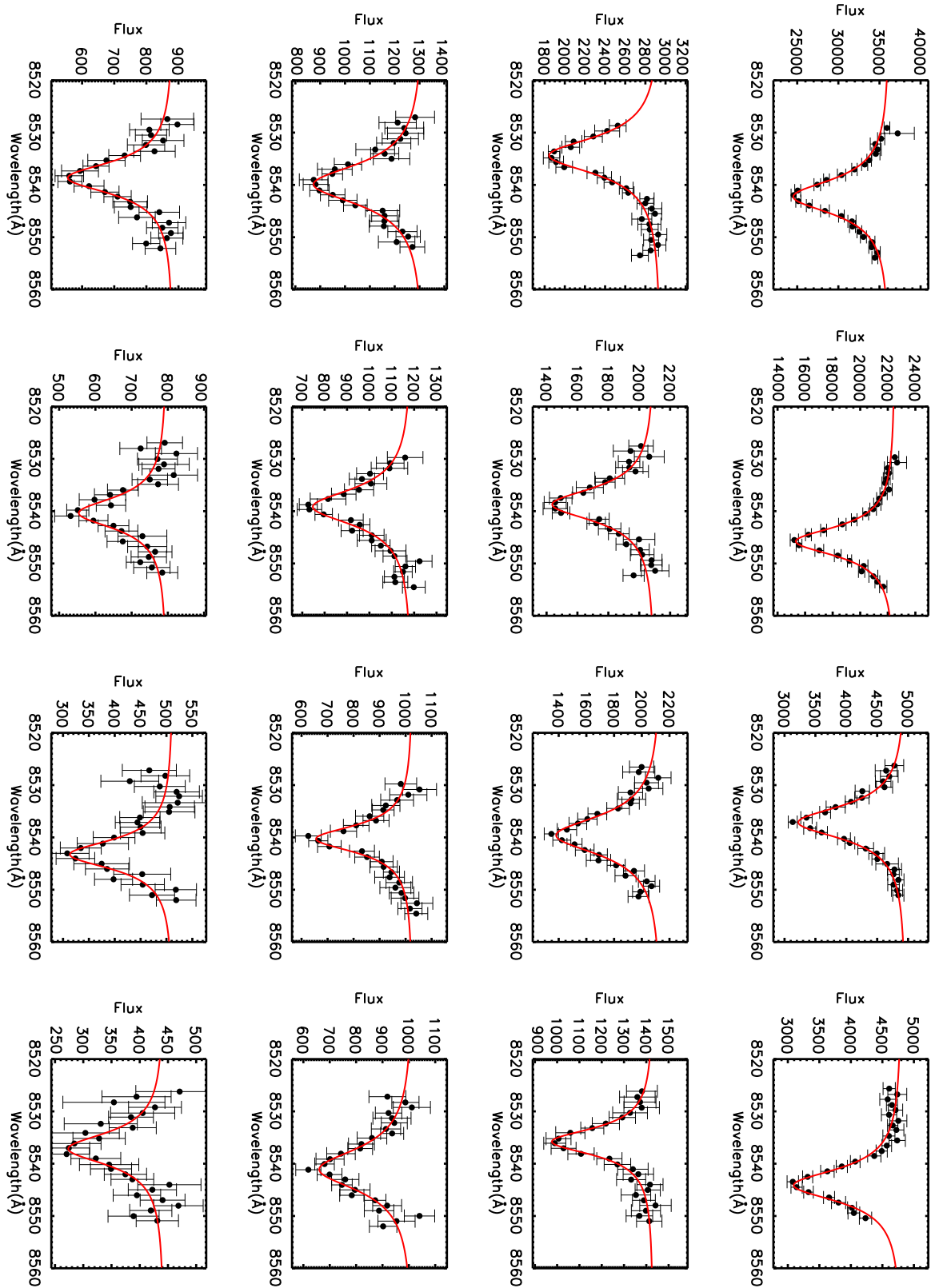}\\
 \caption{A sample of the \ca\ spectra for
 almost the entire brightness range of our sample from I-band magnitude of 11.0 to 16.5.
 Each data point on the absorption line comes from a single image in the data cube.
 The solid line is a Voigt fit to the spectrum.}
 \label{spec}
\end{figure*}
\subsection{Fabry-P\'{e}rot Photometry}
The images in each data cube are overscan corrected, trimmed and bias subtracted using
IRAF\footnote{IRAF is distributed by NOAO, which is operated by AURA, Inc., under a cooperative
agreement with the NSF}. To perform photometry on the individual images we use the DAOPHOT
stand-alone package by \citet{stetson87} that was specially developed to perform crowded field
stellar photometry. The task FIND gives the initial centroids of all the stars in the field. After
performing the initial aperture photometry using the PHOT task, PSF-based photometry is performed
using a quadratically variable PSF, which we found to be especially useful at the edges of the
field. We used about 50 -- 100 stars per image to determine the PSF. The ALLSTAR task then
simultaneously fits all of the stars in the image using the measured PSF to give a star subtracted
image, and fluxes and positions for all stars. Because DAOPHOT performs local sky subtraction, the
fluxes are not affected by the presence of the weak OH sky lines. DAOPHOT includes the ALLFRAME
\citep{stetson94} program that is particularly useful for reducing a data cube. It simultaneously
performs the PSF fitting for all stars in all the images of the cube to derive positions and
magnitudes in a self-consistent manner, making use of the geometric transformations between the
individual images and photometry information for all images. Because our fields are much less
crowded than, for example, a globular cluster, DAOPHOT is able to resolve the stars easily. The
final output of DAOPHOT are the centroids, fluxes (in arbitrary units) and flux uncertainty for
each star.
\subsection{Flux Normalization}
In FP absorption line work, it is critically important to account for changes in flux due to
variations in atmospheric transmission. This can be significant if the images of a spectral scan
were taken over several nights or even in different years, or in non-photometric weather
conditions. To measure the image-to-image normalization, we select about 30 high S/N stars
uniformly covering the FOV, so that there is a range in velocity and wavelength coverage. We
assume that all of these stars have a single dominant absorption line that is well represented by
a Voigt function. We fit a Voigt profile to the spectrum of each star, and at each point calculate
the ratio of the fit to the observed flux. For each image in the scan, we then calculate the mean
and standard deviation of these ratios, using all of the normalization stars for that image. This
ratio is the normalization factor for that image. If an image was taken in average transparency
conditions, then its normalization factor should be 1.0, within the photometric uncertainty; if
the image had lower transparency, the normalization will be greater than unity. We correct the
fluxes for each image in the scan using the individual normalization factors. In order to
determine the final normalization factors, we re-fit the profile to the normalized data, and
iterate the entire process until it converges.

This procedure introduces an additional uncertainty in the flux measurement, which we call the
`normalization uncertainty'. We combine this in quadrature with the DAOPHOT flux uncertainties.
For the highest S/N\footnote{We define the S/N for an individual star as the mean ratio of the
measured flux to the estimated flux uncertainty, averaged over all the points measured in the
star's spectrum.} ($\sim 80$) stars the flux normalization error dominates, while for the majority
of stars the DAOPHOT flux error is more important. Since the different normalization stars have a
range of velocities, and also because of the center-to-edge wavelength gradient in a FP image, an
individual image samples the stars at several different points on the line profile; this avoids
``building in" a particular profile shape through the normalization factors. Under photometric
conditions, the normalization factors were typically $1.00 \pm 0.02$, with uncertainties of $0.002
- 0.003$. In bad conditions, the worst normalizations were $1.5 - 2.0$, with uncertainties of
$\sim 0.01$. We often re-observed at some of the wavelengths taken under poor conditions, and
found excellent agreement between stellar fluxes after normalization.
\subsection{Spectroscopy}
The \ca \ absorption line profiles with the combined uncertainties are fit with a Voigt function
using the Levenberg-Marquardt method of least-squares fitting. The fit gives the central
wavelength, continuum level, line strength, and the Gaussian and Lorentzian widths along with
their measurement uncertainties. We find that the Gaussian contribution to the shape of the
absorption line is negligible for the majority of stars and the line could be fit equally well by
fixing the Gaussian width at $10^{-3}$ \AA, using essentially a Lorentzian profile. This is
because the instrumental profile itself is essentially Lorentzian with a negligible Gaussian core,
and the \ca \ line in these stars is strongly saturated with damping wings from collisional
broadening. Figure \ref{spec} shows a sub-sample of absorption lines covering almost the entire
brightness range for our sample from I-band magnitude of 10.0 to 16.5. The Voigt fit is
represented by the solid line. For the brightest stars the velocity uncertainties can be as low as
2 \kms\, and even for the very faint stars we can measure the absorption line adequately well with
velocity uncertainties of 20 -- 30 \kms.
\section{Radial Velocities}
\begin{figure*}
\centering
\includegraphics[angle=90, scale=0.7]{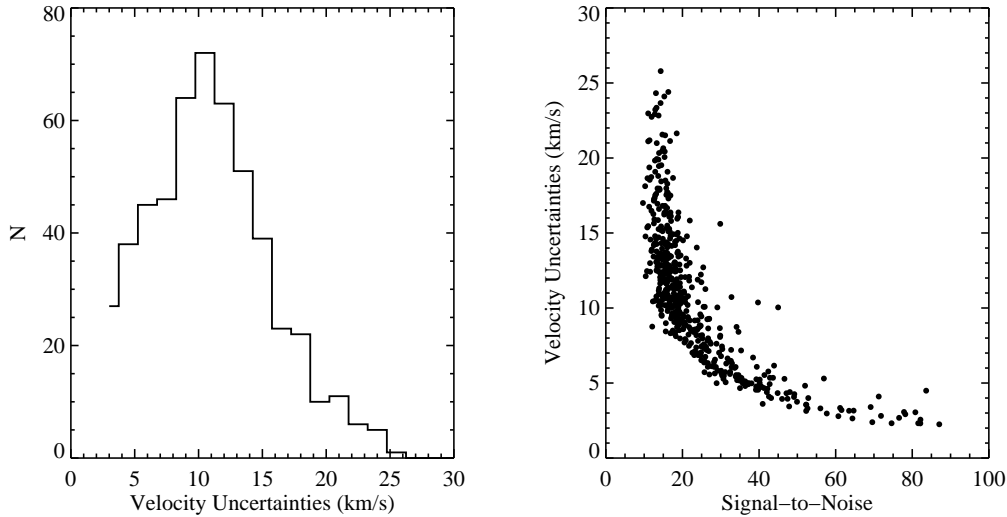}\\
\caption{The distribution of velocity errors for the MM5B-c field (left panel) and velocity
uncertainties as a function of signal-to-noise for the same field (right panel).} \label{verr}
\end{figure*}
Table \ref{data} lists the heliocentric radial velocities and their uncertainties that we measured
for our complete sample of 3360 stars with 1605 and 1037 stars at $l = \pm 5$\degr, respectively,
and 718 in BW. The median velocity error for our sample is $\sim 12$ \kms. The uncertainties are
higher for fainter stars and also in cases where a star has a close or unresolved neighbor. In
Figure \ref{verr} we show the distribution of velocity errors and the error as a function of S/N
for the MM5B-c subfield. The major source of error in the radial velocity measurements comes from
the uncertainties in the flux measurements, which include contributions from DAOPHOT and
normalization errors. In the case of very high S/N stars in the best seeing sub-fields like MM5B-c
and BWC-a (with lower normalization errors) the velocity errors can be as small as 2 -- 5 \kms.

To check if the velocity errors obtained from our Voigt fitter are reasonable, we carried out
Monte Carlo simulations of the data. For each star in our sample, we generated 1000 random
realizations of its spectrum by perturbing the fluxes of each point of the Voigt profile fit by
the corresponding $1-\sigma$ individual flux uncertainty. We then fit a Voigt function to each of
these generated spectra and find that the mean ratio of the error estimate from a single fit to
the standard deviation of the velocities from the Monte Carlo fits is about 0.97 for all the
fields, indicating that the single fit errors from our Voigt fitter code are reliable.

\begin{figure}[t]
\includegraphics[scale=0.60]{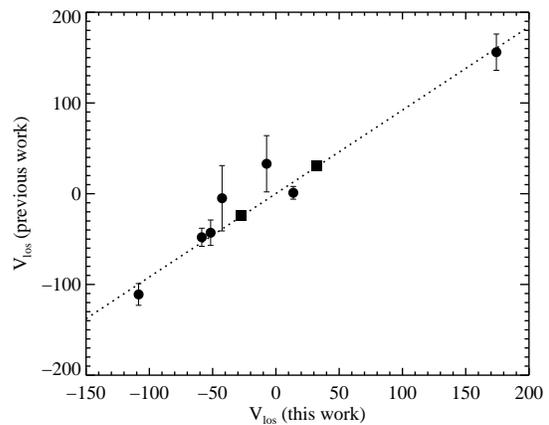}\\
\caption{Radial velocities of 9 stars in common between this work and previous measurements by
\citet{sharples} (squares) and \citet{terndrup} (circles). \label{common}}
\end{figure}

For the BW LOS, there are nine bulge M/K-giant stars in our work that are also present in the
radial velocity samples of \citet{sharples} and \citet{terndrup}. The velocity comparison is shown
in Figure \ref{common}. Our error bars on the horizontal axis are equal to or smaller than the
size of the data points. Our velocities agree very well with the previous measurements, with a
reduced chi-square of 1.02.

\begin{figure*}
 \includegraphics[scale=0.65,angle=90]{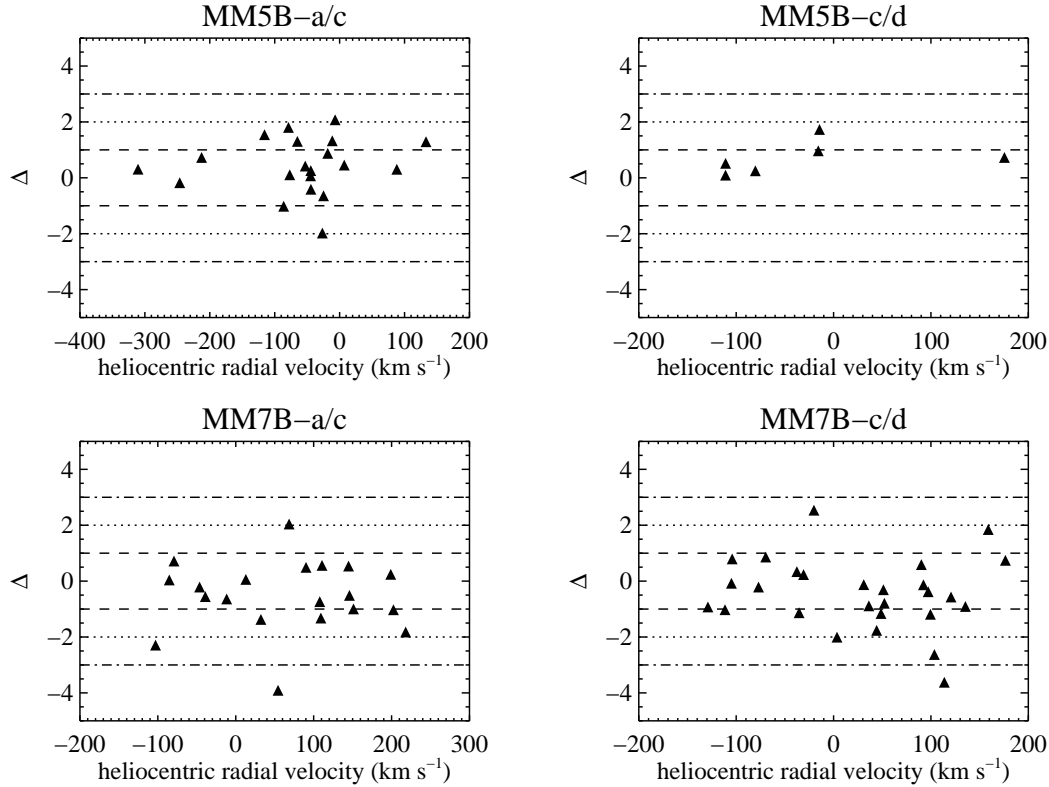}\\
  \caption{Repeated radial velocity measurements of stars in the overlapping regions of
  the sub-fields of MM5B and MM7B.}\label{comm}
\end{figure*}

We also have two measurements (repeated over 1 -- 3 years) of the radial velocities for some stars
in the region where the fields at a given LOS overlap with each other. The comparison between the
velocities of stars in these overlapping regions is shown in Figure \ref{comm} where the vertical
axis shows the velocity difference divided by the $1\sigma$ error bars ($\Delta$). The dashed,
dotted and dot-dashed lines show the $1\sigma$, $2\sigma$ and $3\sigma$ limits respectively. The
stars at the very edge of the field in the overlapping regions may not show up in every image
taken in the scan of that field, so they are not included in this comparison. There is also a 10\%
chance for a foreground disk main sequence star to be a spectroscopic binary which may result in
velocity variations of $\lesssim 30$ \kms\,\citep{DM91}. We find three such examples of foreground
stars with large velocity differences in the overlapping regions, and these three stars are
omitted from Figure \ref{comm}. The comparison for the MM5B overlapping regions has a reduced
chi-square of $\lesssim 1$. There is one additional star (not foreground) in each of the two
overlapping regions of the MM7B fields that has a deviation of more than $3 \sigma$, inflating the
reduced chi-square for these fields. We don't have a clear explanation for the deviations in these
two cases except that they could be spectroscopic binaries in the bulge. The reduced chi-square
for all the stars in the MM7B fields is 1.6; excluding the two outliers reduces chi-square to
1.06.

For most of the stars in the overlapping region we list the weighted mean velocity and errors in
Table \ref{data}. For the faintest stars from the overlapping regions, we use the velocity from
only the field with the best image quality, where the fit to the absorption line was significantly
better than the other field. The comparison with previous work, and in the overlapping regions,
indicates that our velocity measurements are repeatable and reliable. The velocity uncertainty of
less than 30 \kms\ is small compared to the velocity dispersion of the stars in the bar, and will
have no effect on our conclusions.

\section{Photometry and Astrometry}
We use photometric information to distinguish between the foreground disk stars, bulge red giants
and bulge RCGs. Since our observed fields are coincident with the OGLE survey, we have precision
(1 -- 2\%) V and I band photometry and
astrometry\footnote{http://ogledb.astrouw.edu.pl/$\sim$ogle/photdb/} \citep{database}.
Unfortunately, the fields MM7B-d and a quarter of MM7B-c fell outside the OGLE field. Photometric
observations for these fields were obtained with the 1-m telescope at the South African
Astronomical Observatory (SAAO). Reduction of these data using the IRAF photometry package gives V
and I band magnitudes with an accuracy of 5\% -- 10\%.

There are some stars in our sample that have no OGLE or SAAO photometry. For these stars we use
the FP measurement of the continuum level of the \ca\ absorption line to estimate I-band
magnitudes. To check the accuracy of these FP I-band magnitudes we use all the stars in the MM7Bc
field where we have both OGLE and SAAO photometry. We determine the zero point of the FP I-band
magnitudes from the OGLE photometry for these stars. We compare them to the OGLE and the SAAO
magnitudes to find that they are surprisingly consistent with differences $\lesssim 10\%$. These
FP I-band magnitudes are used to exclude foreground stars as discussed in section 6. Comparison
with the much deeper OGLE survey shows that our sample of RCGs in the bar is essentially complete.
Table \ref{data} lists the astrometry and photometry. The photometry source (OGLE, SAAO or FP) is
indicated in the last column. The astrometry for all fields comes from OGLE except for the field
MM7B-d where the source is the Digitized Sky Survey.
\section{Selections from the Color-Magnitude Diagram}
\begin{figure}[t]
 \includegraphics[scale=0.85, angle=90]{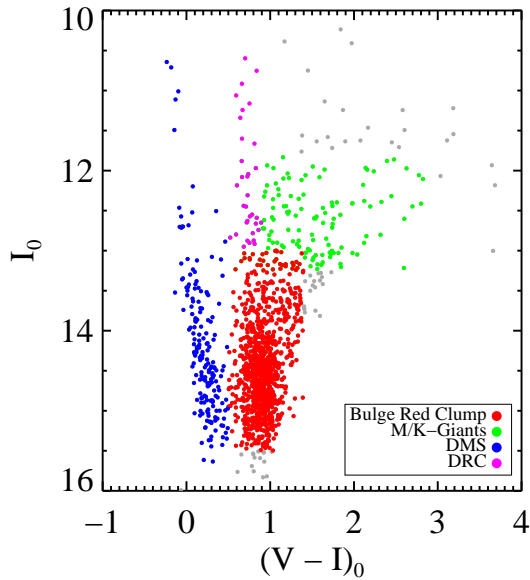}
 \caption{Extinction-corrected color-Magnitude diagram for the MM5B LOS shows a distinct red clump (red dots).
 The disk main sequence and the M/K -- giants are shown in blue and green dots respectively.
 We measure the radial velocity of every star in the diagram.}
 \label{mm5bselectfun}
\end{figure}

The color-magnitude diagram (CMD) is shown in Figure \ref{mm5bselectfun} for the MM5B field with
1605 stars. We chose to show the CMD for this particular LOS since it has the largest number of
stars and also has OGLE photometry for all them. The CMDs for the other two LOS look almost
identical. \citet{pac94} discuss various parts and features of the BW CMD using the photometry of
about $3 \times 10^{5}$ stars from the OGLE project. We discuss in brief these features and their
selection limits in the $(I_0, (V - I)_0)$ plane to be used in our analysis.

The first feature is a distinct red clump, marked by the red points. These RCGs are very useful
candidates for studying the kinematics of the bar. They are basically metal-rich counterparts of
the horizontal branch stars burning helium in their cores and hydrogen in their envelopes. Their
intrinsic luminosity distribution is very narrow and has a weak dependence on metallicity and age,
making them excellent distance indicators \citep{pac98}. They are bright and numerous in the inner
galaxy and are located in the region of CMD that is least contaminated by disk stars
\citep{stanek94}.

Our selection limits for RCGs towards all three LOS are $(I_0, (V-I)_0) = (13.0 - 15.5, 0.5 -
1.4)$, corrected for the extinction ($A_I$) and reddening ($E_{V-I}$) values from \citet{sumi}.
This I-band limit corresponds to a distance range of about 5.5 -- 13 kpc for intrinsic red clump
magnitude $M_I = -0.28 \pm 0.20$ \citep{pac98}. \citet{stanek97} used a slightly different
selection criteria for OGLE RCGs where they define extinction independent I-band magnitudes. The
effect of these two different selection methods on the overall kinematics of the RCGs is
negligible, with changes in the mean radial velocity and dispersion of about 0.75 \kms\, and 1.15
\kms\, respectively. The latter selection also gives 2\% fewer stars at $l = \pm 5$\degr and 10\%
fewer stars in BW. We have also investigated the effect of varying the brighter end of the I-band
selection limit from 13.0 -- 13.6 magnitudes, which produced negligible changes to the mean
velocity and dispersion ($\sim 0.5$ \kms). The effects of different selection criteria for RCGs
will be more significant when modeling the data (subject of paper III), where it will be
investigated in greater detail.

About 210 stars at $(l,b) = (\pm 5.0,-3.5)$\degr and 39 in BW have only I-band photometry. Without
color information it is not possible to distinguish them from the foreground disk stars (blue
points in Figure \ref{mm5bselectfun}). If we apply only I-band cuts then the maximum difference in
the kinematics between including or excluding these stars without color information is about 1.3
\kms\, in mean velocity and 0.6 \kms\, in dispersion. The 2MASS point source catalogue has J, H
and K band photometry for our lines of sight. We found J-K colors for 60\% of these stars with no
V-I color information. The J-K colors confirmed that after applying the I-band or distance cut
about 96\% of these stars were RCGs. By analyzing the field MM7B-c, where 99\% of the sample has
complete V and I band photometry, we find that the contamination due to lack of color information
is $\sim 14\%$, after applying the I-band or distance cut. This amounts to contamination by only
11 stars at $l = \pm 5$\degr and 2 stars in BW. These numbers are negligible compared to the
overall sample size and will have little effect on the kinematics. Thus we include these stars
(with just the I-band limits) in our final sample that will be used to measure bar kinematics,
which now has a total of 1193 RCGs at $(l,b) = (-5.0, -3.5)$\degr, 738 at $(l,b) = (5.5,
-3.5)$\degr and 557 in BW.

The region marked by the green points at roughly $(I_0, (V-I)_0) = (11.8 - 13.25, 0.5 - 3.0)$
contains the bulge M/K-giants \citep{sharples,terndrup}. This is the same M-giant population used
in the recent radial velocity survey by \citet{rich07}. In section 7.2 we use them to make a
direct comparison with that study. This sample, measured simultaneously from the same data cube,
allows us to compare the kinematics of two different stellar populations in the bar/bulge region.

The feature marked by the blue points at $(I_0, (V-I)_0) = (10.0 - 15.7, -0.3 - 0.5)$ is formed by
the disk main sequence (DMS) stars. The stars in this feature are thought to be associated with
the Sagittarius spiral arm located at $\sim 2$ kpc from the Sun \citep{pac94, ng96}. We have
radial velocities of about 350 stars in this region over three LOS. This feature and its possible
association with the Sagittarius arm is discussed in section 7.3.
\begin{figure*}
 \includegraphics[scale=0.7,angle=90]{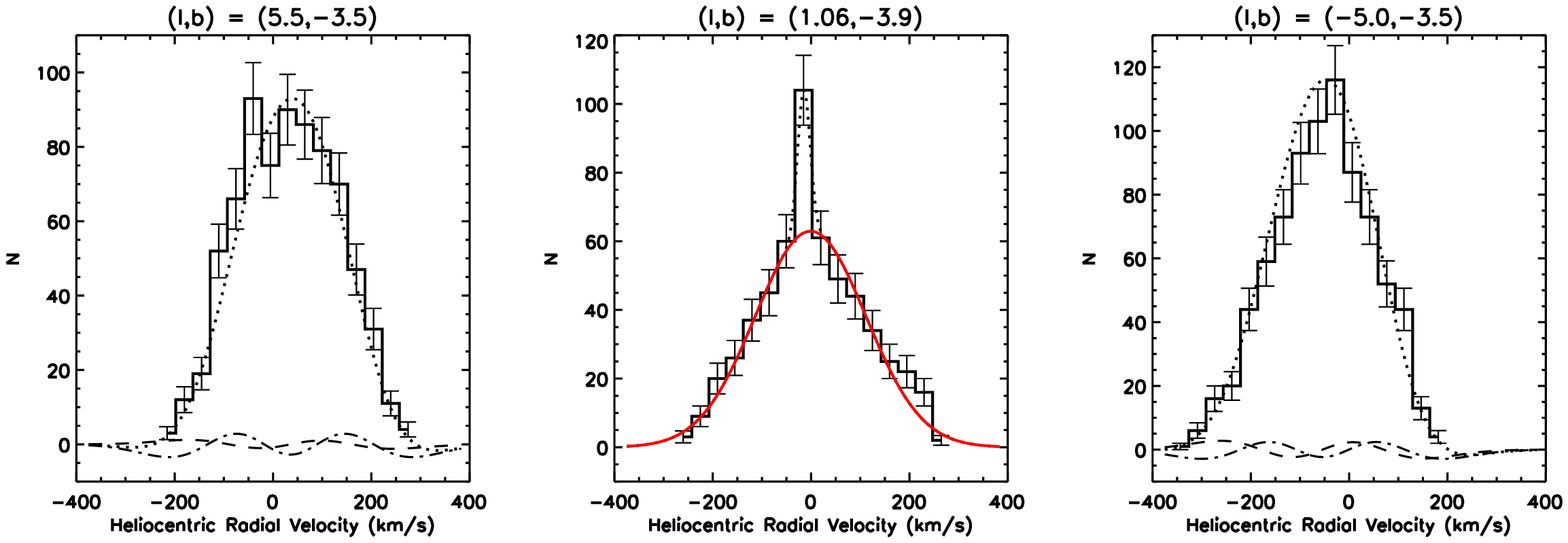}\\
 \caption{Stellar radial velocity distribution for three LOS in the bar.
 Right and Left panel: the dotted line is the Gauss-Hermite fit with $h_3$ and $h_4$ moments shown
 by dashed and dot-dashed lines respectively. Middle panel: the dotted line is double
 Gaussian with a narrow component for the globular cluster NGC 6522 and the solid line is the broad
 component (only) for the Galactic bar/bulge.}
 \label{vhist}
\end{figure*}

Finally, the feature at $(I_0,(V-I)_0) = (10.0 - 13.0, 0.5 - 0.9)$\footnote{Note that there is an
overlap in the selection limits between other features and the M-giant sample. We chose the
selection of the M-giants to be as close as possible to that used in previous investigations, in
order to make kinematic comparisons.} on the CMD marked by the magenta points consists of the disk
red clump giants spread over a distance of about 1.0 -- 4.0 kpc. We had just enough stars in our
sample along this LOS to detect this feature and measure its kinematics. The FP imaging
spectroscopy has allowed us to measure simultaneously the kinematics of all these features,
providing a powerful and reliable way to look for differences between the various stellar
populations in a consistent manner.
\section{Kinematics}
\subsection{Kinematics of Red Clump Giants in the Bar: Measurement of $h_3$ and $h_4$}
The large size of our sample allows us to measure, for the first time, the higher moments of the
velocity distribution along our three LOS. We fit a Gauss-Hermite series to the velocity
distribution \citep{gerhard93,gh93} of RCGs to extract the four velocity moments -- mean line of
sight velocity ($<V_{\textrm{los}}>$), line-of-sight velocity dispersion
($\sigma_{\textrm{los}}$), asymmetry of the distribution ($h_3$), and flatness of the distribution
($h_4$)  -- using a maximum likelihood estimator (MLE). An advantage of using the MLE is that it
corrects for the individual velocity errors that can broaden the width of the distribution. We use
the MLE code developed by Glenn Van de Ven (private communication; see \citet{glenn06} for more
details) to fit a Gauss-Hermite series to our velocity distributions and to determine the velocity
moments. The code gives a robust measurement of the uncertainties of each moment in two ways. The
bootstrap method randomly draws from the observed velocity distribution with replacement
\citep[see Sect. 15.6 of][]{press92}; this gives the maximum possible errors of the parameters.
The Gaussian randomization (GR) method randomly draws velocities from an intrinsic Gaussian
distribution with a given mean and dispersion that is broadened by velocity errors randomly drawn
from the observed velocity error distribution \citep[see Appendix B of ][]{glenn06}. GR gives
reliable estimates if the individual errors are not underestimated, which we believe is the case
with our analysis. Finally, the code uses the biweight statistic to measure just the first two
velocity moments ($<V_{\textrm{los}}>$ and $\sigma_{\textrm{los}}$) and their uncertainties; these
are the appropriate measures to compare with previous investigations that had too few stars to
reliably determine the higher moments.

MM5B-a and MM5B-c have identical kinematics and can be combined into a single field MM5B-(a+c).
MM5B-d is at the same LOS and overlaps with the MM5B-c sub-field but has a mean velocity that is
offset by 13 \kms\ (i.e. about a $2\sigma$ deviation). The dispersions in all three sub-fields
however agree within the $1\sigma$ errors. For this reason we measure the velocity moments at $l =
-5.0$\degr\ using both MM5B-(a+c) as well as the three sub-fields combined MM5B-(a+c+d). There are
no significant differences in the kinematics of sub-fields MM7B-a, MM7B-c and MM7B-d, so they are
combined into a single field MM7B-(a+c+d). The velocity distributions for MM5B-(a+c) and
MM7B-(a+c+d) are shown in the left and right panel of Figure \ref{vhist}. The dotted line is the
Gauss-Hermite fit with the $h_3$ and $h_4$ moments represented by the dashed and dot-dashed lines
respectively. Table \ref{kine} lists the velocity moments along with their uncertainties. Our
large sample size allows us to measure $<V_{\textrm{los}}>$ and $\sigma_{\textrm{los}}$ with high
accuracy. As expected, in Galactocentric coordinates $<V_{\textrm{los}}>$ has equal amplitude of
about $\pm 63$ \kms\ at $l = \pm 5$\degr, respectively. The $h_4$ moment is significantly non-zero
while the $h_3$ is consistent with zero. Our high S/N velocity distributions show no sign of cold
kinematic features that could be related to disrupted satellites, as suggested by \citet{rich07}.
\begin{figure*}
 \includegraphics[scale=0.8,angle=90]{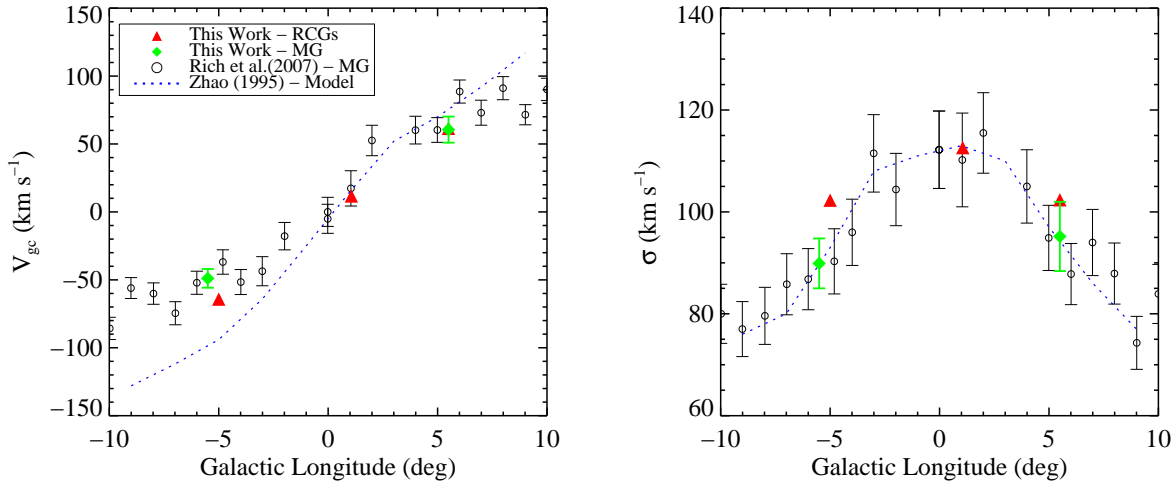}\\
 \caption{Comparison of the kinematics (in galactocentric frame) of RCGs (red triangles) and M-giants (green diamonds) from our sample to
 \citet{rich07} data (open circles) and \citet{zhao96} model (blue dashed line)}
 \label{comp}
\end{figure*}

The analysis of the BW velocity distribution (middle panel) is slightly complicated by the
presence of stars from the bulge globular cluster NGC 6522, the outskirts of which extend into the
southwest edge of our field (see Figure \ref{stellarfield}). The spike at the center of the BW
velocity distribution (shown in Figure \ref{vhist} middle panel) is caused by the stars that are
part of this cluster. To extract kinematics we fit a double Gaussian (broad for the bulge and
narrow for NGC 6522) to this distribution. The narrow Gaussian has a mean velocity of $-14.67 \pm
3.77$ \kms \ and dispersion of $7.9 \pm 2.82$ \kms \ ; consistent with previous measurements for
NGC 6522 of $-10.3 \pm 1.6$ \kms \ and $7.3^{+3.5}_{-2.0}$ \kms \, respectively \citep{dubath97}.
The dotted line in Figure \ref{vhist} shows the double gaussian fit to the distribution and the
solid line is the broad component representing the bulge velocity distribution. We also fit a
Gauss-Hermite series to this distribution and list the values in Table \ref{kine}, although this
measurement is most probably affected by the presence of the spike at the center. Since this is
the most heavily studied LOS, there are several measurements of the kinematics in BW for
comparison. Our values of the mean line-of-sight velocity and dispersion are in excellent
agreement with all previous measurements of the kinematics in BW such as \citet{mould83}
($<V_{\textrm{los}}> = -10 \pm 19$ \kms, $\sigma_{\textrm{los}} = 113 \pm 11$ \kms),
\citet{sharples} ($<V_{\textrm{los}}> = 4 \pm 8$ \kms, $\sigma_{\textrm{los}} = 113^{+6}_{-5}$
\kms), \citet{ rich90} ($<V_{\textrm{los}}> = -13 \pm 12$ \kms, $\sigma_{\textrm{los}} = 111 \pm
8$ \kms), \citet{terndrup} ($<V_{\textrm{los}}> = -8 \pm 6$ \kms, $\sigma_{\textrm{los}} = 110 \pm
10$ \kms) and \citet{rich07} ($<V_{\textrm{los}}> = -1.1\pm 12.9$ \kms, $\sigma_{\textrm{los}} =
110.2 \pm 9.1$ \kms).

\subsection{Kinematics of M-giants}
Our measurements of the kinematics of M-giants (in the heliocentric coordinate system) are listed
in Table \ref{kine2}. We do not tabulate the measurements for BW since the globular cluster
M-giants dominate the already small sample, and are difficult to separate from the bulge
population, making the measurement unreliable. In Figure \ref{comp} we compare the kinematics of
the bulge RCGs and M-giants from our sample with that of \citet{rich07} M-giants and
\citet{zhao96} (Zhao hereafter) model. We find that the radial velocities of the RCGs and M-giants
stars agree well with each other and with the \citet{rich07} measurements. However the LOS
dispersion for our M-giant sample (green diamonds) is systematically lower than the RCGs (red
triangles), and is consistent with the velocity dispersion of \citet{rich07} M-giants (black
circles). Note that the higher dispersion of the RCGs in our sample is not because they are
fainter and thus have somewhat larger velocity uncertainties. The MLE method we use removes the
effects of the uncertainties from the estimate of the dispersion. In any event, to produce the
difference between the M-giant and the RCG dispersion would require addition in quadrature of 40
-- 50 \kms\, compared to the uncertainties of 10 -- 25 \kms.

The blue dashed line in Figure \ref{comp} is the prediction from Zhao's model. The bar in this
model rotates faster than the observed rotation curve at greater longitudes. According to Zhao and
\citet{rich07} this rapid rotation is due to a lack of retrograde orbits in the model. But adding
retrograde orbits will increase the $\sigma_{\textrm{los}}$ of the model, which may then agree
with the dispersion of RCGs rather than that of the M-giants.

\begin{table}[t]
\caption{Kinematics of bar M-giants and disk main sequence stars}
\begin{tabular}{cccc}
 \tableline\tableline
 LOS & $<V_{\mathrm{los}}>$ & $\sigma_{\textrm{los}}$ & N\\
  & ($\mathrm{km\, s^{-1}}$) & ($\mathrm{km\, s^{-1}})$ & \\
 \tableline
 \multicolumn{4}{c}{M-giants}\\
 \tableline
 (-5.0, -3.5) & $-37.2 \pm 6.9$ & $89.9 \pm 4.9$ & 171\\
 (5.5, -3.5) & $29.9 \pm 9.6$ & $95.2 \pm 6.8$ & 98\\
 \tableline
 \multicolumn{4}{c}{DMS}\\
 \tableline
 (-5.0, -3.5) & $-17.8\pm 4.0$ & $45.9 \pm 2.8$ & 134\\
 (1.06, -3.9) & $ -3.1\pm 4.8$ & $41.3 \pm 3.4$ & 75\\
 (5.5, -3.5) & $0.9 \pm 4.3$ & $45.7\pm 3.1$ & 109\\
 \tableline
\end{tabular}\label{kine2}
\end{table}

Is there really a difference in dispersion between the M-giants and RCGs? One possibility is that
the M-giant sample may have more foreground contamination, which could cause the observed
decrement in the dispersion. However, our M-giant selection criteria are virtually the same as
those of \citet{rich07}, who argue that contamination is negligible. If the difference is real,
then perhaps there is an age difference between the RCGs and the M-giants, with RCG's being older.
Another possibility is that the M-giants represent multiple dynamical populations. But at this
point we have no means to distinguish between these or other hypotheses. We plan to obtain more
lines-of-sight in the bar using the FP system \citep{rangwala08} on the 10-m Southern African
Large Telescope (SALT) \citep{buckley03} to explore this discrepancy further.
\subsection{Kinematics of Disk Stars}
Our measurements of the kinematics of the DMS stars (blue dots in figure \ref{mm5bselectfun}) are
listed in Table \ref{kine2}. Earlier studies of the Baade's window CMD \citep{terndrup, pac94,
ng96} found an excess of stars within 2.5 kpc of the Sun. They associated this increased stellar
density (by a factor $\sim 2$) of DMS stars with the location of the Sagittarius spiral arm at a
distance of $\sim 2$ kpc. The simulations performed by \citet{ng96} showed this density enhanced
feature to consist of some very young stars (0.1 -- 2.0 Gyr) that are identified with the recent
spiral arm population superimposed on a larger population of young and intermediate age (2.0 --
7.0 Gyr) stars identified with some past spiral arm population. We find the
$\sigma_{\textrm{los}}$ to be the same ($\sim$ 45 \kms) towards all three LOS. According to the
relationship between velocity dispersion and distance from the Galactic center \citep{freeman89},
our dispersion corresponds to a position within 3.5 kpc of the Sun. This suggests that the
majority of the stars in this feature may belong to the Sagittarius spiral arm population.
\begin{figure}[t]
\includegraphics[scale=0.7]{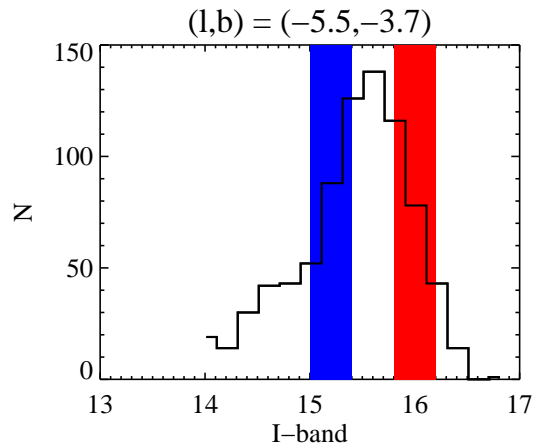}
\caption{I-band apparent magnitude distribution of the red clump giants in MM5B field. The two
bands show the sample limits for stars on the near and far sides of the bar.\label{idist}}
\end{figure}

The dispersion of the disk red clump feature (Magenta points on the CMD) is about 53 \kms,
consistent with the disk kinematics.
\section{Radial Streaming Motion}
Previous observations have established that the near end of the Galactic bar is in the first
Galactic quadrant, i.e., at positive longitudes. Since the stars forming the bar stream in the
same sense as the Galactic rotation, for this bar orientation the stars on the near and far side
of the bar will have radial velocity components towards and away from us, respectively. The radial
velocity difference between these approaching and receding stellar streams depends on the
orientation of the bar, among other things. Since RCGs are good distance indicators with a small
dispersion in intrinsic luminosity, they can be localized in space and used to measure precisely
the velocity shift between the two streams (\citet{mao02}; MP02 hereafter). This measurement will
provide an additional strong constraint on the dynamical models of the Galactic bar, especially
its orientation angle.
\begin{figure*}
  \includegraphics[scale=0.7, angle=90]{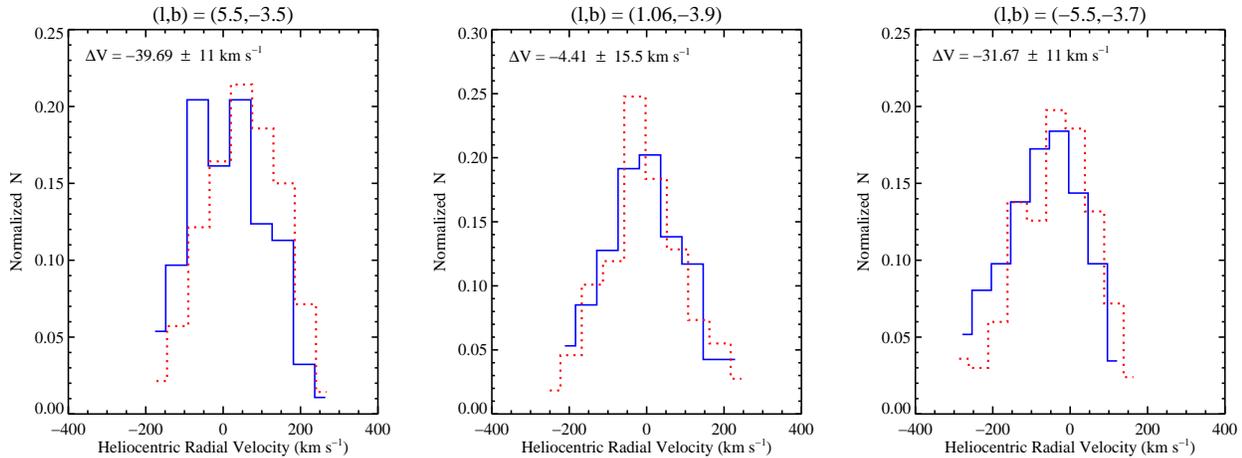}\\
  \caption{The radial velocity distribution for the stars in the near
  and far side samples, shown by the solid and dotted histograms respectively.}\label{stream}
\end{figure*}

MP02 discuss the optimal selection of the bright and faint samples to isolate these stellar
streams. We adapt their criteria for the characteristics of our data set, selecting RCGs 0.2 to
0.6 magnitudes brighter and fainter than the peak in the observed luminosity distribution for each
field to define the bright (front side) and faint (back side), respectively. Figure \ref{idist}
illustrates the two samples for the l = -5.0\degr\ field, and the details of these bright and
faint samples for the three LOS are listed in Table \ref{streamdata}. Figure \ref{stream} shows
the velocity distribution of the near and far side samples. We detect a clear velocity shift of
$\sim 35$ \kms\ in each of $l \simeq \pm 5$\degr\ LOS. The velocity shift in BW is much smaller
and is consistent with zero. This may be because in BW we are seeing an older spheroidal bulge
population superimposed on the bar population, masking this signal. The higher velocity dispersion
in BW also indicates that there are higher random motions compared to the streaming motions. Our
observed shifts of 35 \kms\ are much greater than any possible projection effects arising from the
Galactic rotation and the location of the near and far samples (1 -- 2 \kms). This suggests that
there are strong non-circular streaming motions associated with the bar. MP02 predicted the
difference of the LOS component of the streaming motions between the bright and faint sample to be
$\sim 33$ \kms\, based on the E2 model of \citet{dwek95}. This predicted shift is consistent with
our measurements in the $l = \pm 5$\degr\ fields but not in BW. MP02 also discussed the effects of
red giant contamination on these shifts and concluded that their presence in the RCG sample could
significantly dilute the magnitude of the velocity shifts, so that the actual shifts could be
larger than those that we measure.

\citet{deguchi01} \citep[see also][]{deguchi} used SiO masers in the bar to measure the velocity
shifts. They find an average radial velocity difference of $-21 \pm 27$ \kms\ for sources with
distances less than 7 kpc compared to those at distances greater than 7 kpc, essentially
independent of longitude. While their results agree with our measurements, they suffer from a
large (50\%) uncertainty in distance and large statistical errors in kinematics (due to a smaller
sample size).

\begin{figure*}
  \includegraphics[scale=0.7,angle=90]{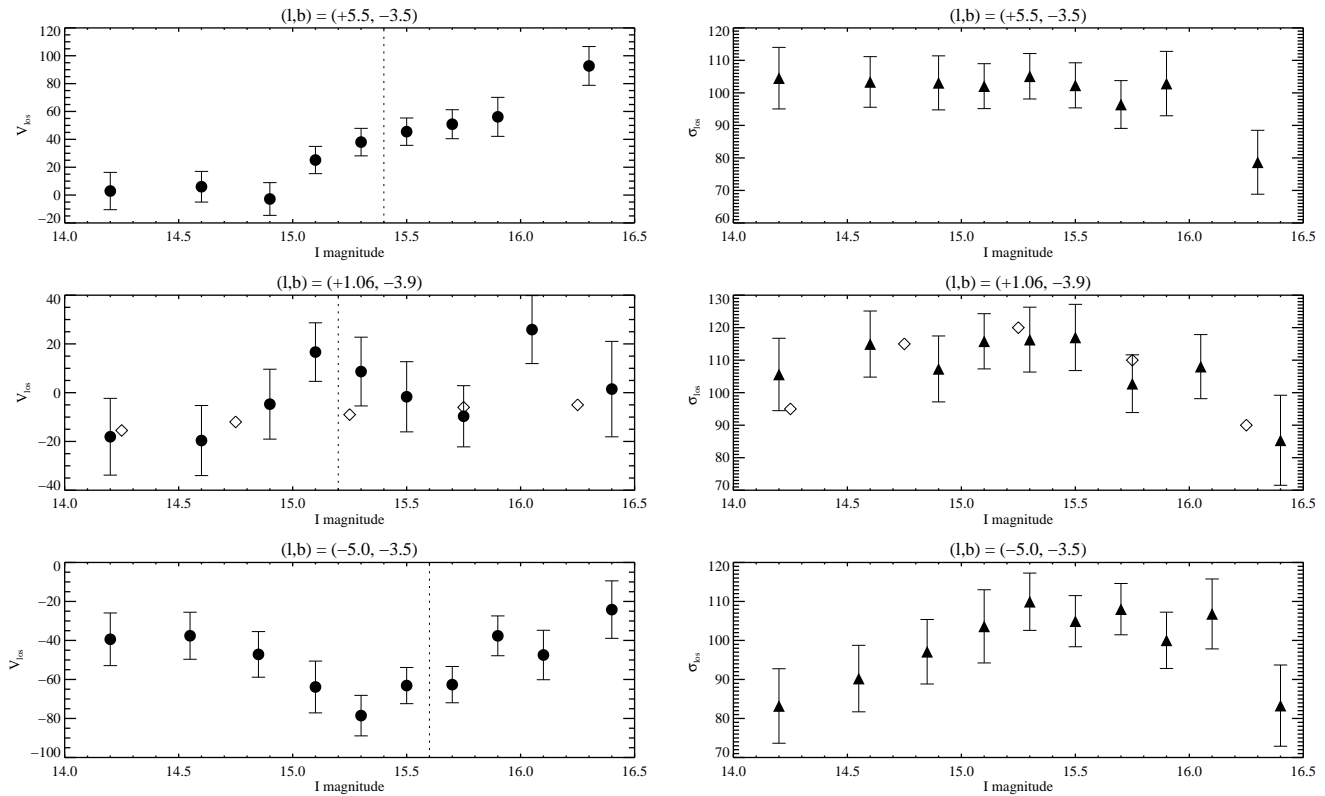}\\
  \caption{Line-of-sight radial velocity and dispersion as a function of I-band
magnitude (or distance) of the bulge red clump giants for three LOS in the Galactic bar. The
vertical dashed line shows the position of the peak magnitude about which the bright and faint
samples were measured. Note that these are apparent magnitude not corrected for extinction. The
open diamonds show \citet{hafner}'s dynamical model.}\label{vimag}
\end{figure*}

As pointed out by \citet{deguchi01} most of the dynamical models of the inner galaxy do not
predict or even discuss the change in average radial velocity with distance. The dynamical model
of \citet{hafner} was one of the first to discuss this in some detail. In their Figures 9 and 10
they plot line-of-sight and proper motion kinematics as a function of distance for four windows,
one of which is BW. We have a large enough sample of RCGs to plot LOS average velocity and
dispersion as a function of distance in the inner galaxy, which we show in Figure \ref{vimag}.
Although we cannot make an exact comparison with Hafner's predictions for BW since they use a
generic selection function for K giants, our data are nonetheless in excellent agreement with
their model (open diamonds in Figure \ref{vimag}).

\section{Equivalent Widths}
Table \ref{data} lists the equivalent widths and their uncertainties for our complete sample of
3360 stars. We measured the equivalent widths of \ca \ lines in our sample by integrating the
Voigt fit over all wavelengths. Our measurements of mean equivalent width and dispersion for the
bulge RCGs are listed in Table \ref{ew}. The mean equivalent width ($<\mathrm{EW}>$) and its
dispersion decrease on either side of BW similar to the LOS velocity dispersion. Because the \ca \
line is a good proxy for metallicity \citep{az88}, the gradient in the mean equivalent width may
imply a gradient in metallicity \citep{frogel88, minniti95} along the Galactic bar. We also find
that the mean EW of the DMS stars is about 25\% lower than the bulge RCGs, which is consistent
with a bulge that is more metal rich than the solar neighborhood. This has been the consensus of
many previous studies in BW, for example \citet{rich88}. The factors affecting the measurement of
FP equivalent widths and inferring [Fe/H] using the calcium infrared triplet method is addressed
in detail in Paper II. For recent work on the metallicity of the Galactic bulge see
\citet{fulbright06, minniti08}.
\section{Discussion and Conclusions}
\begin{table}[t]
\caption{Mean equivalent width and dispersion for the \ca\ line in the bar RCGs}\label{ew}
\begin{tabular}{cccc}
 \tableline\tableline
 LOS & $<\mathrm{EW}>$ & $\sigma_{\mathrm{EW}}$ & N\\
  & \AA & \AA & \\
 \tableline
(-5.0,-3.5) & $3.68 \pm 0.04$ & $0.46\pm 0.04$ & 804\\
(1.1,-3.9) & $3.86 \pm 0.05$ & $0.68\pm 0.04$ & 557\\
(5.5,-3.5) & $3.28 \pm 0.04$ & $0.48\pm 0.04$ & 738\\
\tableline
\end{tabular}
\end{table}

In this paper we have presented measurements of the \ca\ absorption line using Fabry-P\'{e}rot
imaging spectroscopy. This work shows the strength and reliability of this technique in measuring
radial velocities for a large sample of stars with I-band magnitudes ranging from 10.0 -- 16.5.
Past investigations have mostly used red giants to study the bar's stellar kinematics. Our work is
the first to measure the kinematics of red clump giants, which are more numerous in the inner
galaxy than the M and K giants. The main focus of this paper is to present the data and
preliminary interpretations, but full interpretation will only be possible with extensive modeling
that we will present in Paper III.

We obtained radial velocities for three LOS in the MW bar that include two offset positions at $l
\simeq \pm 5$\degr and the central field of BW. The large sample size enabled high precision
measurements of the first four moments of the velocity distributions for the first time. The
symmetric moment $h_4$ is significantly non-zero, and its negative value indicates that the
distributions are flat-topped rather than peaked. This seems consistent with a model of the bar
with stars in elongated orbits forming two streams at different mean radial velocities, broadening
and flattening the total distribution. Because our sample of bar RCGs is essentially complete, the
continuity equation suggests that the number of stars in the two streams should be the same. Table
\ref{streamdata} shows that this is the case within the Poisson noise. We then expect no asymmetry
in the velocity distribution function, and our measures of $h_3$ are indeed consistent with zero.

When we divide our sample into bright and faint samples we detect a clear kinematic signature of
streaming motions about the bar. The stellar streams have velocity difference of about 35~\kms\ at
$l = \pm 5$\degr. There is no indication of these streams in BW, perhaps because they are masked
by a spheroidal bulge component. The larger velocity dispersion in BW is consistent with this
interpretation. Our large sample size allows us to take the first step towards measuring the LOS
kinematics as a function of apparent magnitude (or distance) using the RCGs. As \citet{hafner}
points out, these measurements can provide powerful constraints on dynamical models of the inner
Galaxy. With larger data sets it will be possible to detect the transition from disk to bar, and
measure the bar's thickness.

The technique of FP imaging spectroscopy has allowed us to measure the radial velocities samples
that are an order of magnitude larger (at individual LOS) than previous studies. This large sample
allows us to determine the kinematics of various stellar populations: bar RCGs, bar M-giants and
the disk main sequence stars. Our measurements show that the RCG and M-giant populations have the
same $<V_{\textrm{los}}>$ but their dispersions differ by about 10 \kms. The kinematics of the DMS
population suggests that most of these stars belong to the Sagittarius spiral arm.

In addition to the kinematic measurement, we have used the FP spectra to measure the \ca\ line
strengths. We can use these results to measure the metallicities of the individual stars.
Metallicity distributions for several LOS in the bar/bulge can provide strong constraints on the
chemical evolution models that can determine the star formation rate, initial mass function, and
formation time scale \citep[e.g.][]{matteucci99,zoccali07} for the inner Galaxy. Our measurements
indicate a gradient in mean and dispersion of equivalent width as a function of Galactic
longitude, and a clear difference in the mean line strengths between the disk and bulge/bar. The
metallicity measurements will be the subject of Paper II.

Motivated by the success of this project, we plan to use the FP system on the 10-m class SALT to
obtain a much more extensive determination of the kinematics and metallicity of stars in the inner
Galaxy. SALT's much greater aperture and larger FOV will enable us to measure $\gtrsim 2000$ stars
along a single LOS in an hour of observing time. We will investigate at least 10 LOS along the
major axis of the bar, obtaining 15000 -- 20000 stellar spectra. This data set will provide the
basis for models of the structure and chemical evolution of the Galactic bar and bulge. We can
also investigate the long bar suggested by recent Spitzer observations \citep{benjamin08} and
search for hypervelocity stars. This planned increase by another order of magnitude in the size of
the data sample will revolutionize the studies of the inner Galaxy.

\acknowledgements Bohdan Paczynski was an early and enthusiastic collaborator on this project, and
provided insight and inspiration that contributed greatly to its success. Jerry Sellwood has
provided the theoretical foundation for the program and cheerfully tutored our understanding of
bar dynamics. Tad Pryor was our DAOPHOT expert and statistical guru.  Rutgers undergraduates
Robert Polocz and Brian Schnitzer assisted with data reduction. The director and staff of CTIO
provided their usual outstanding level of support for the observations, especially when amazingly
bad weather and recalcitrant equipment threatened to scuttle the project. We thank the director of
SAAO for allocating observing time and Dr. John Menzies for obtaining the additional photometry
observations. Victor Debattista has provided invaluable help in developing dynamical models using
these data. This research has made use of the SIMBAD database, operated at CDS, Strasbourg,
France. This project has been supported by the National Science Foundation through grant
AST0507323.



\begin{thebibliography}{}

\bibitem[{{Alcock} {et~al.}(2000){Alcock}, {Allsman}, {Alves}, {Axelrod},
  {Becker}, {Bennett}, {Cook}, {Drake}, {Freeman}, {Geha}, {Griest}, {Lehner},
  {Marshall}, {Minniti}, {Nelson}, {Peterson}, {Popowski}, {Pratt}, {Quinn},
  {Stubbs}, {Sutherland}, {Tomaney}, {Vandehei}, \& {Welch}}]{alcock00}
{Alcock}, C., {Allsman}, R.~A., {Alves}, D.~R., {Axelrod}, T.~S., {Becker},
  A.~C., {Bennett}, D.~P., {Cook}, K.~H., {Drake}, A.~J., {Freeman}, K.~C.,
  {Geha}, M., {Griest}, K., {Lehner}, M.~J., {Marshall}, S.~L., {Minniti}, D.,
  {Nelson}, C.~A., {Peterson}, B.~A., {Popowski}, P., {Pratt}, M.~R., {Quinn},
  P.~J., {Stubbs}, C.~W., {Sutherland}, W., {Tomaney}, A.~B., {Vandehei}, T.,
  \& {Welch}, D.~L. 2000, \apj, 541, 734

\bibitem[{{Armandroff} \& {Zinn}(1988)}]{az88}
{Armandroff}, T.~E., \& {Zinn}, R. 1988, \aj, 96, 92

\bibitem[{{Benjamin}(2008)}]{benjamin08}
{Benjamin}, R.~A. 2008, in Bulletin of the American Astronomical Society,
  Vol.~40, Bulletin of the American Astronomical Society, 266--+

\bibitem[{{Benjamin} {et~al.}(2005){Benjamin}, {Churchwell}, {Babler},
  {Indebetouw}, {Meade}, {Whitney}, {Watson}, {Wolfire}, {Wolff}, {Ignace},
  {Bania}, {Bracker}, {Clemens}, {Chomiuk}, {Cohen}, {Dickey}, {Jackson},
  {Kobulnicky}, {Mercer}, {Mathis}, {Stolovy}, \& {Uzpen}}]{benjamin05}
{Benjamin}, R.~A., {Churchwell}, E., {Babler}, B.~L., {Indebetouw}, R.,
  {Meade}, M.~R., {Whitney}, B.~A., {Watson}, C., {Wolfire}, M.~G., {Wolff},
  M.~J., {Ignace}, R., {Bania}, T.~M., {Bracker}, S., {Clemens}, D.~P.,
  {Chomiuk}, L., {Cohen}, M., {Dickey}, J.~M., {Jackson}, J.~M., {Kobulnicky},
  H.~A., {Mercer}, E.~P., {Mathis}, J.~S., {Stolovy}, S.~R., \& {Uzpen}, B.
  2005, \apjl, 630, L149

\bibitem[{{Binney} {et~al.}(1991){Binney}, {Gerhard}, {Stark}, {Bally}, \&
  {Uchida}}]{binney91}
{Binney}, J., {Gerhard}, O.~E., {Stark}, A.~A., {Bally}, J., \& {Uchida}, K.~I.
  1991, \mnras, 252, 210

\bibitem[{{Bissantz} {et~al.}(2004){Bissantz}, {Debattista}, \&
  {Gerhard}}]{bissantz04}
{Bissantz}, N., {Debattista}, V.~P., \& {Gerhard}, O. 2004, \apjl, 601, L155

\bibitem[{{Bissantz} \& {Gerhard}(2002)}]{bissantz02}
{Bissantz}, N., \& {Gerhard}, O. 2002, \mnras, 330, 591

\bibitem[{{Blitz} \& {Spergel}(1991)}]{blitz91}
{Blitz}, L., \& {Spergel}, D.~N. 1991, \apj, 379, 631

\bibitem[{{Buckley} {et~al.}(2003){Buckley}, {Hearnshaw}, {Nordsieck}, \&
  {O'Donoghue}}]{buckley03}
{Buckley}, D.~A.~H., {Hearnshaw}, J.~B., {Nordsieck}, K.~H., \& {O'Donoghue},
  D. 2003, in Presented at the Society of Photo-Optical Instrumentation
  Engineers (SPIE) Conference, Vol. 4834, Discoveries and Research Prospects
  from 6- to 10-Meter-Class Telescopes II. Edited by Guhathakurta, Puragra.
  Proceedings of the SPIE, Volume 4834, pp. 264-275 (2003)., ed.
  P.~{Guhathakurta}, 264--275

\bibitem[{{Cabrera-Lavers} {et~al.}(2008){Cabrera-Lavers},
  {Gonzalez-Fernandez}, {Garzon}, {Hammersley}, \& {Lopez-Corredoira}}]{cl08}
{Cabrera-Lavers}, A., {Gonzalez-Fernandez}, C., {Garzon}, F., {Hammersley},
  P.~L., \& {Lopez-Corredoira}, M. 2008, ArXiv e-prints, 809

\bibitem[{{Clarkson} {et~al.}(2008){Clarkson}, {Sahu}, {Anderson}, {Smith},
  {Brown}, {Rich}, {Casertano}, {Bond}, {Livio}, {Minniti}, {Panagia},
  {Renzini}, {Valenti}, \& {Zoccali}}]{clarkson08}
{Clarkson}, W., {Sahu}, K., {Anderson}, J., {Smith}, T.~E., {Brown}, T.~M.,
  {Rich}, R.~M., {Casertano}, S., {Bond}, H.~E., {Livio}, M., {Minniti}, D.,
  {Panagia}, N., {Renzini}, A., {Valenti}, J., \& {Zoccali}, M. 2008, \apj,
  684, 1110

\bibitem[{{de Vaucouleurs}(1964)}]{devac64}
{de Vaucouleurs}, G. 1964, in IAU Symposium, Vol.~20, The Galaxy and the
  Magellanic Clouds, ed. F.~J. {Kerr}, 195--+

\bibitem[{{Debattista} \& {Sellwood}(2000)}]{victor00}
{Debattista}, V.~P., \& {Sellwood}, J.~A. 2000, \apj, 543, 704

\bibitem[{{Debattista} \& {Williams}(2004)}]{victor04}
{Debattista}, V.~P., \& {Williams}, T.~B. 2004, \apj, 605, 714

\bibitem[{{Deguchi}(2001)}]{deguchi}
{Deguchi}, S. 2001, in Astronomical Society of the Pacific Conference Series,
  Vol. 228, Dynamics of Star Clusters and the Milky Way, ed. S.~{Deiters},
  B.~{Fuchs}, A.~{Just}, R.~{Spurzem}, \& R.~{Wielen}, 404--+

\bibitem[{{Deguchi} {et~al.}(2001){Deguchi}, {Fujii}, {Matsumoto}, {Nakashima},
  \& {Wood}}]{deguchi01}
{Deguchi}, S., {Fujii}, T., {Matsumoto}, S., {Nakashima}, J.-I., \& {Wood},
  P.~R. 2001, \pasj, 53, 293

\bibitem[{{Dubath} {et~al.}(1997){Dubath}, {Meylan}, \& {Mayor}}]{dubath97}
{Dubath}, P., {Meylan}, G., \& {Mayor}, M. 1997, \aap, 324, 505

\bibitem[{{Duquennoy} \& {Mayor}(1991)}]{DM91}
{Duquennoy}, A., \& {Mayor}, M. 1991, \aap, 248, 485

\bibitem[{{Dwek} {et~al.}(1995){Dwek}, {Arendt}, {Hauser}, {Kelsall}, {Lisse},
  {Moseley}, {Silverberg}, {Sodroski}, \& {Weiland}}]{dwek95}
{Dwek}, E., {Arendt}, R.~G., {Hauser}, M.~G., {Kelsall}, T., {Lisse}, C.~M.,
  {Moseley}, S.~H., {Silverberg}, R.~F., {Sodroski}, T.~J., \& {Weiland}, J.~L.
  1995, \apj, 445, 716

\bibitem[{{Freeman}(1996)}]{freeman96}
{Freeman}, K.~C. 1996, in Astronomical Society of the Pacific Conference
  Series, Vol.~91, IAU Colloq. 157: Barred Galaxies, ed. R.~{Buta}, D.~A.
  {Crocker}, \& B.~G. {Elmegreen}, 1--+

\bibitem[{{Freudenreich}(1998)}]{cobe98}
{Freudenreich}, H.~T. 1998, \apj, 492, 495

\bibitem[{{Frogel}(1988)}]{frogel88}
{Frogel}, J.~A. 1988, \araa, 26, 51

\bibitem[{{Fulbright} {et~al.}(2006){Fulbright}, {McWilliam}, \&
  {Rich}}]{fulbright06}
{Fulbright}, J.~P., {McWilliam}, A., \& {Rich}, R.~M. 2006, \apj, 636, 821

\bibitem[{{Gebhardt} {et~al.}(1994){Gebhardt}, {Pryor}, {Williams}, \&
  {Hesser}}]{gebhardt94}
{Gebhardt}, K., {Pryor}, C., {Williams}, T.~B., \& {Hesser}, J.~E. 1994, \aj,
  107, 2067

\bibitem[{{Gerhard}(1993)}]{gerhard93}
{Gerhard}, O.~E. 1993, \mnras, 265, 213

\bibitem[{{Gerhard}(1999)}]{gerhard99}
{Gerhard}, O.~E. 1999, in Astronomical Society of the Pacific Conference
  Series, Vol. 182, Galaxy Dynamics - A Rutgers Symposium, ed. D.~R. {Merritt},
  M.~{Valluri}, \& J.~A. {Sellwood}, 307--+

\bibitem[{{Ghavamian} {et~al.}(2003){Ghavamian}, {Rakowski}, {Hughes}, \&
  {Williams}}]{ghavamian03}
{Ghavamian}, P., {Rakowski}, C.~E., {Hughes}, J.~P., \& {Williams}, T.~B. 2003,
  \apj, 590, 833

\bibitem[{{Habing} {et~al.}(2006){Habing}, {Sevenster}, {Messineo}, {van de
  Ven}, \& {Kuijken}}]{habing}
{Habing}, H.~J., {Sevenster}, M.~N., {Messineo}, M., {van de Ven}, G., \&
  {Kuijken}, K. 2006, \aap, 458, 151

\bibitem[{{H{\"a}fner} {et~al.}(2000){H{\"a}fner}, {Evans}, {Dehnen}, \&
  {Binney}}]{hafner}
{H{\"a}fner}, R., {Evans}, N.~W., {Dehnen}, W., \& {Binney}, J. 2000, \mnras,
  314, 433

\bibitem[{{Hartigan} {et~al.}(2000){Hartigan}, {Morse}, {Palunas}, {Bally}, \&
  {Devine}}]{hartigan00}
{Hartigan}, P., {Morse}, J., {Palunas}, P., {Bally}, J., \& {Devine}, D. 2000,
  \aj, 119, 1872

\bibitem[{{Howard} {et~al.}(2008){Howard}, {Rich}, {Reitzel}, {Koch}, {De
  Propris}, \& {Zhao}}]{howard08}
{Howard}, C.~D., {Rich}, R.~M., {Reitzel}, D.~B., {Koch}, A., {De Propris}, R.,
  \& {Zhao}, H. 2008, ArXiv e-prints, 807

\bibitem[{{Lewis} \& {Freeman}(1989)}]{freeman89}
{Lewis}, J.~R., \& {Freeman}, K.~C. 1989, \aj, 97, 139

\bibitem[{{Liszt} \& {Burton}(1980)}]{liszt80}
{Liszt}, H.~S., \& {Burton}, W.~B. 1980, \apj, 236, 779

\bibitem[{{Mao} \& {Paczy{\'n}ski}(2002)}]{mao02}
{Mao}, S., \& {Paczy{\'n}ski}, B. 2002, \mnras, 337, 895

\bibitem[{{Matteucci} \& {Romano}(1999)}]{matteucci99}
{Matteucci}, F., \& {Romano}, D. 1999, \apss, 265, 311

\bibitem[{{Minniti} {et~al.}(1995){Minniti}, {Olszewski}, {Liebert}, {White},
  {Hill}, \& {Irwin}}]{minniti95}
{Minniti}, D., {Olszewski}, E.~W., {Liebert}, J., {White}, S.~D.~M., {Hill},
  J.~M., \& {Irwin}, M.~J. 1995, \mnras, 277, 1293

\bibitem[{{Minniti} {et~al.}(1992){Minniti}, {White}, {Olszewski}, \&
  {Hill}}]{minniti92}
{Minniti}, D., {White}, S.~D.~M., {Olszewski}, E.~W., \& {Hill}, J.~M. 1992,
  \apjl, 393, L47

\bibitem[{{Minniti} \& {Zoccali}(2008)}]{minniti08}
{Minniti}, D., \& {Zoccali}, M. 2008, 245, 323

\bibitem[{{Mould}(1983)}]{mould83}
{Mould}, J.~R. 1983, \apj, 266, 255

\bibitem[{{Ng} {et~al.}(1996){Ng}, {Bertelli}, {Chiosi}, \& {Bressan}}]{ng96}
{Ng}, Y.~K., {Bertelli}, G., {Chiosi}, C., \& {Bressan}, A. 1996, \aap, 310,
  771

\bibitem[{{Osterbrock} \& {Martel}(1992)}]{osterbrock}
{Osterbrock}, D.~E., \& {Martel}, A. 1992, \pasp, 104, 76

\bibitem[{{Paczynski} \& {Stanek}(1998)}]{pac98}
{Paczynski}, B., \& {Stanek}, K.~Z. 1998, \apjl, 494, L219+

\bibitem[{{Paczynski} {et~al.}(1994){Paczynski}, {Stanek}, {Udalski},
  {Szymanski}, {Kaluzny}, {Kubiak}, \& {Mateo}}]{pac94}
{Paczynski}, B., {Stanek}, K.~Z., {Udalski}, A., {Szymanski}, M., {Kaluzny},
  J., {Kubiak}, M., \& {Mateo}, M. 1994, \aj, 107, 2060

\bibitem[{{Palunas} \& {Williams}(2000)}]{palunas}
{Palunas}, P., \& {Williams}, T.~B. 2000, \aj, 120, 2884

\bibitem[{{Press} {et~al.}(1992){Press}, {Teukolsky}, {Vetterling}, \&
  {Flannery}}]{press92}
{Press}, W.~H., {Teukolsky}, S.~A., {Vetterling}, W.~T., \& {Flannery}, B.~P.
  1992, {Numerical recipes in FORTRAN. The art of scientific computing}
  (Cambridge: University Press, |c1992, 2nd ed.)

\bibitem[{{Rangwala} {et~al.}(2008){Rangwala}, {Williams}, {Pietraszewski}, \&
  {Joseph}}]{rangwala08}
{Rangwala}, N., {Williams}, T.~B., {Pietraszewski}, C., \& {Joseph}, C.~L.
  2008, \aj, 135, 1825

\bibitem[{{Rattenbury} {et~al.}(2007{\natexlab{a}}){Rattenbury}, {Mao},
  {Debattista}, {Sumi}, {Gerhard}, \& {de Lorenzi}}]{rattenbury07a}
{Rattenbury}, N.~J., {Mao}, S., {Debattista}, V.~P., {Sumi}, T., {Gerhard}, O.,
  \& {de Lorenzi}, F. 2007{\natexlab{a}}, \mnras, 378, 1165

\bibitem[{{Rattenbury} {et~al.}(2007{\natexlab{b}}){Rattenbury}, {Mao}, {Sumi},
  \& {Smith}}]{rattenbury07b}
{Rattenbury}, N.~J., {Mao}, S., {Sumi}, T., \& {Smith}, M.~C.
  2007{\natexlab{b}}, \mnras, 378, 1064

\bibitem[{{Rich}(1988)}]{rich88}
{Rich}, R.~M. 1988, \aj, 95, 828

\bibitem[{{Rich}(1990)}]{rich90}
---. 1990, \apj, 362, 604

\bibitem[{{Rich} {et~al.}(2007){Rich}, {Reitzel}, {Howard}, \& {Zhao}}]{rich07}
{Rich}, R.~M., {Reitzel}, D.~B., {Howard}, C.~D., \& {Zhao}, H. 2007, \apjl,
  658, L29

\bibitem[{{Sellwood} \& {Wilkinson}(1993)}]{sellwood93}
{Sellwood}, J.~A., \& {Wilkinson}, A. 1993, Reports of Progress in Physics, 56,
  173

\bibitem[{{Sharples} {et~al.}(1990){Sharples}, {Walker}, \&
  {Cropper}}]{sharples}
{Sharples}, R., {Walker}, A., \& {Cropper}, M. 1990, \mnras, 246, 54

\bibitem[{{Sluis} \& {Williams}(2006)}]{sluis06}
{Sluis}, A.~P.~N., \& {Williams}, T.~B. 2006, \aj, 131, 2089

\bibitem[{{Spaenhauer} {et~al.}(1992){Spaenhauer}, {Jones}, \&
  {Whitford}}]{span}
{Spaenhauer}, A., {Jones}, B.~F., \& {Whitford}, A.~E. 1992, \aj, 103, 297

\bibitem[{{Stanek} {et~al.}(1994){Stanek}, {Mateo}, {Udalski}, {Szymanski},
  {Kaluzny}, \& {Kubiak}}]{stanek94}
{Stanek}, K.~Z., {Mateo}, M., {Udalski}, A., {Szymanski}, M., {Kaluzny}, J., \&
  {Kubiak}, M. 1994, \apjl, 429, L73

\bibitem[{{Stanek} {et~al.}(1997){Stanek}, {Udalski}, {Szymanski}, {Kaluzny},
  {Kubiak}, {Mateo}, \& {Krzeminski}}]{stanek97}
{Stanek}, K.~Z., {Udalski}, A., {Szymanski}, M., {Kaluzny}, J., {Kubiak}, M.,
  {Mateo}, M., \& {Krzeminski}, W. 1997, \apj, 477, 163

\bibitem[{{Stetson}(1987)}]{stetson87}
{Stetson}, P.~B. 1987, \pasp, 99, 191

\bibitem[{{Stetson}(1994)}]{stetson94}
---. 1994, \pasp, 106, 250

\bibitem[{{Sumi}(2004)}]{sumi}
{Sumi}, T. 2004, \mnras, 349, 193

\bibitem[{{Sumi} {et~al.}(2004){Sumi}, {Wu}, {Udalski}, {Szyma{\'n}ski},
  {Kubiak}, {Pietrzy{\'n}ski}, {Soszy{\'n}ski}, {Wo{\'z}niak},
  {{\.Z}ebru{\'n}}, {Szewczyk}, \& {Wyrzykowski}}]{sumi04}
{Sumi}, T., {Wu}, X., {Udalski}, A., {Szyma{\'n}ski}, M., {Kubiak}, M.,
  {Pietrzy{\'n}ski}, G., {Soszy{\'n}ski}, I., {Wo{\'z}niak}, P.,
  {{\.Z}ebru{\'n}}, K., {Szewczyk}, O., \& {Wyrzykowski}, {\L}. 2004, \mnras,
  348, 1439

\bibitem[{{Szymanski}(2005)}]{database}
{Szymanski}, M.~K. 2005, Acta Astronomica, 55, 43

\bibitem[{{Terndrup} {et~al.}(1995){Terndrup}, {Sadler}, \& {Rich}}]{terndrup}
{Terndrup}, D.~M., {Sadler}, E.~M., \& {Rich}, R.~M. 1995, \aj, 110, 1774

\bibitem[{{Udalski} {et~al.}(1992){Udalski}, {Szymanski}, {Kaluzny}, {Kubiak},
  \& {Mateo}}]{udalski}
{Udalski}, A., {Szymanski}, M., {Kaluzny}, J., {Kubiak}, M., \& {Mateo}, M.
  1992, Acta Astronomica, 42, 253

\bibitem[{{van Albada} \& {Sanders}(1982)}]{sanders82}
{van Albada}, T.~S., \& {Sanders}, R.~H. 1982, \mnras, 201, 303

\bibitem[{{van de Ven} {et~al.}(2006){van de Ven}, {van den Bosch}, {Verolme},
  \& {de Zeeuw}}]{glenn06}
{van de Ven}, G., {van den Bosch}, R.~C.~E., {Verolme}, E.~K., \& {de Zeeuw},
  P.~T. 2006, \aap, 445, 513

\bibitem[{{van der Marel} \& {Franx}(1993)}]{gh93}
{van der Marel}, R.~P., \& {Franx}, M. 1993, \apj, 407, 525

\bibitem[{{Weiner} \& {Sellwood}(1999)}]{weiner99}
{Weiner}, B.~J., \& {Sellwood}, J.~A. 1999, \apj, 524, 112

\bibitem[{{Z{\'a}nmar S{\'a}nchez} {et~al.}(2008){Z{\'a}nmar S{\'a}nchez},
  {Sellwood}, {Weiner}, \& {Williams}}]{zanmar08}
{Z{\'a}nmar S{\'a}nchez}, R., {Sellwood}, J.~A., {Weiner}, B.~J., \&
  {Williams}, T.~B. 2008, \apj, 674, 797

\bibitem[{{Zhao}(1996)}]{zhao96}
{Zhao}, H.~S. 1996, \mnras, 283, 149

\bibitem[{{Zoccali} {et~al.}(2007){Zoccali}, {Lecureur}, {Barbuy}, {Hill},
  {Renzini}, {Minniti}, {Momany}, {G{\'o}mez}, \& {Ortolani}}]{zoccali07}
{Zoccali}, M., {Lecureur}, A., {Barbuy}, B., {Hill}, V., {Renzini}, A.,
  {Minniti}, D., {Momany}, Y., {G{\'o}mez}, A., \& {Ortolani}, S. 2007, 241, 73

\end{thebibliography}

\clearpage

\clearpage
\begin{deluxetable}{ccccrcccccc}
\tabletypesize{\scriptsize} \tablecaption{Radial Velocities and Equivalent Widths}
\tablehead{\colhead{ID}& \colhead{OGLEID} & \colhead{R.A.} & \colhead{Dec.} & \colhead{$V$} &
\colhead{$V_{\textrm{err}}$} & \colhead{EW} & \colhead{EW$_{\textrm{err}}$} & \colhead{I} & \colhead{V-I} & \colhead{Source}\\
\colhead{} & \colhead{}& \colhead{j2000} & \colhead{j2000} & \colhead{km $\textrm{s}^{-1}$} &
\colhead{km $\textrm{s}^{-1}$} & \colhead{\AA} & \colhead{\AA} & \colhead{} & \colhead{} &
\colhead{}}

\startdata
BWCa-1  &        256217  &     18:03:35.16  &     -29:58:12.3  &     -4.49  &    1.60  &      4.64  &      0.17  &    11.560  &     2.117  &  o  \\
BWCa-2  &        256200  &     18:03:31.91  &     -30:00:28.3  &    -27.71  &    2.55  &      2.12  &      0.12  &    11.523  &     3.960  &  o  \\
BWCa-3  &        256193  &     18:03:31.67  &     -30:00:43.4  &    -44.86  &    1.91  &      4.55  &      0.19  &    11.730  &     2.522  &  o  \\
BWCa-4  &        402256  &     18:03:43.52  &     -30:01:46.1  &    -17.77  &    1.70  &      4.34  &      0.15  &    11.733  &     1.358  &  o  \\
BWCa-5  &        256201  &     18:03:30.78  &     -30:00:27.9  &     92.33  &    2.41  &      3.79  &      0.18  &    12.241  &     2.702  &  o  \\
BWCa-6  &        256199  &     18:03:32.87  &     -30:00:31.4  &      9.84  &    3.60  &      2.42  &      0.20  &    12.098  &     0.710  &  o  \\
BWCa-8  &        402258  &     18:03:45.27  &     -30:01:32.5  &    -74.90  &    3.48  &      2.94  &      0.22  &    12.214  &     3.876  &  o  \\
BWCa-9  &          --  &     18:03:36.34  &     -30:01:50.2  &    -20.10  &    3.03  &      3.80  &      0.23  &    12.020  &  --  &  f  \\
BWCa-11  &        244433  &     18:03:34.54  &     -30:01:37.6  &     32.20  &    3.41  &      1.20  &      0.11  &    12.229  &     4.566  &  o  \\
BWCa-12  &        412692  &     18:03:38.98  &     -29:58:25.9  &    -38.04  &    2.72  &      5.65  &      0.31  &    12.215  &     2.595  &  o  \\
BWCa-13  &        423223  &     18:03:41.96  &     -29:57:49.1  &    -70.86  &    2.80  &      4.07  &      0.25  &    12.289  &     2.120  &  o  \\
BWCa-15  &        412682  &     18:03:38.18  &     -30:00:51.2  &    -16.60  &    2.91  &      4.07  &      0.23  &    12.467  &     1.912  &  o  \\
BWCa-16  &        412681  &     18:03:38.24  &     -30:00:59.7  &     67.05  &    2.41  &      4.64  &      0.23  &    12.476  &     2.346  &  o  \\
BWCa-19  &        412690  &     18:03:36.96  &     -29:58:47.1  &    -58.58  &    2.45  &      6.04  &      0.31  &    12.483  &     2.094  &  o  \\
BWCa-20  &        256196  &     18:03:33.66  &     -30:00:36.1  &    -24.73  &    2.47  &      3.46  &      0.20  &    12.579  &     1.894  &  o  \\
BWCa-21  &        412685  &     18:03:42.64  &     -30:00:06.5  &    131.70  &    3.01  &      4.69  &      0.32  &    12.493  &     2.927  &  o  \\
BWCa-22  &        402257  &     18:03:37.71  &     -30:01:34.0  &    -15.85  &    2.90  &      3.79  &      0.22  &    12.637  &     2.040  &  o  \\
BWCa-23  &        412683  &     18:03:47.96  &     -30:00:38.1  &    -17.18  &    3.30  &      4.40  &      0.32  &    12.627  &     2.948  &  o  \\
BWCa-27  &        412719  &     18:03:39.69  &     -29:58:47.5  &   -108.48  &    3.33  &      6.48  &      0.35  &    12.787  &     2.576  &  o  \\
BWCa-28  &        412708  &     18:03:49.05  &     -29:59:35.0  &     -7.34  &    2.95  &      5.23  &      0.35  &    12.905  &     2.475  &  o  \\
BWCa-29  &        256222  &     18:03:34.70  &     -30:01:15.7  &    -18.07  &    3.08  &      3.48  &      0.23  &    12.923  &     1.883  &  o  \\
BWCa-30  &        412702  &     18:03:37.64  &     -30:00:26.0  &    -45.72  &    6.02  &      3.66  &      0.41  &    12.877  &     0.925  &  o  \\
BWCa-32  &        256236  &     18:03:34.96  &     -29:59:48.6  &    -76.35  &    5.27  &      4.78  &      0.45  &    12.982  &     3.493  &  o  \\
BWCa-33  &        412727  &     18:03:45.61  &     -29:58:12.0  &   -132.32  &    3.95  &      4.81  &      0.40  &    13.174  &     2.579  &  o  \\
BWCa-34  &        412711  &     18:03:41.51  &     -29:59:20.3  &   -134.86  &    3.59  &      4.75  &      0.26  &    13.081  &     2.852  &  o  \\
BWCa-35  &        256235  &     18:03:30.56  &     -29:59:51.0  &     44.72  &    3.22  &      4.07  &      0.29  &    13.173  &     2.307  &  o  \\
BWCa-36  &        256238  &     18:03:32.78  &     -29:59:39.1  &    143.70  &    2.71  &      4.15  &      0.27  &    13.132  &     2.736  &  o  \\
BWCa-38  &        412728  &     18:03:42.44  &     -29:58:10.6  &   -168.91  &    3.82  &      5.29  &      0.34  &    13.205  &     2.625  &  o  \\
BWCa-39  &        412712  &     18:03:47.60  &     -29:59:15.4  &    -65.47  &    3.57  &      5.50  &      0.38  &    13.223  &     2.865  &  o  \\
BWCa-40  &        412718  &     18:03:38.12  &     -29:58:58.3  &      3.29  &    3.78  &      5.35  &      0.39  &    13.150  &     2.474  &  o  \\
BWCa-42  &        412732  &     18:03:37.09  &     -29:57:56.3  &      4.77  &    3.54  &      2.35  &      0.22  &    13.133  &     1.571  &  o  \\
BWCa-43  &        412724  &     18:03:39.04  &     -29:58:28.0  &   -153.07  &    4.33  &      5.49  &      0.39  &    13.173  &     1.917  &  o  \\
BWCa-44  &        412704  &     18:03:38.05  &     -30:00:10.2  &    -61.60  &    4.52  &      3.80  &      0.27  &    13.141  &     1.419  &  o  \\
BWCa-45  &        412716  &     18:03:47.48  &     -29:59:06.4  &    110.99  &    2.89  &      4.14  &      0.28  &    13.309  &     2.530  &  o  \\
BWCa-46  &        412699  &     18:03:46.37  &     -30:00:47.9  &   -130.64  &    3.93  &      4.99  &      0.36  &    13.333  &     2.210  &  o  \\
BWCa-47  &        412720  &     18:03:45.08  &     -29:58:46.3  &     89.38  &    3.35  &      4.34  &      0.31  &    13.284  &     2.242  &  o  \\
BWCa-48  &        402288  &     18:03:40.62  &     -30:01:46.0  &    -10.80  &    4.33  &      3.03  &      0.30  &    13.377  &     1.751  &  o  \\
BWCa-49  &        412729  &     18:03:39.95  &     -29:58:07.8  &   -102.93  &    4.34  &      4.77  &      0.37  &    13.295  &     1.956  &  o  \\
BWCa-50  &        256226  &     18:03:36.45  &     -30:00:39.8  &    -13.72  &    3.74  &      5.04  &      0.39  &    13.311  &     2.713  &  o  \\
\enddata
 \tablecomments{Table \ref{data} is published in its entirety in the electronic edition of the
{\it Astrophysical Journal}.  A portion is shown here for guidance regarding its form and
content.\label{data}} \tablenotetext{a}{Source of Photometry: `o' for OGLE, `s' for SAAO and `f'
for fabry-p\'{e}rot}
\end{deluxetable}

\begin{deluxetable}{lcccc}
\tabletypesize{\scriptsize} \tablecaption{Heliocentric kinematics of the red clump giants for
three LOS in the Milky Way bar.\label{kine}}
 \tablehead{
 \colhead{Method} & \colhead{$<V_{\textrm{los}}>$} & \colhead{$\sigma_{\textrm{los}}$} & \colhead{$h_3$} & \colhead{$h_4$}}
 \startdata
 \cutinhead{MM5B-(a+c); $(l,b) = (-5.0, -3.5)$; $N_{\textrm{stars}} = 804$}
 Likelihood & $-54.15 \pm 3.66\, (\pm 6.14)$ & $101.72 \pm 0.66\, (\pm 0.80)$ & $-0.030 \pm 0.017\, (\pm 0.034)$
  & $-0.032 \pm 0.008\, (\pm 0.017)$\\
 Biweight  & $-52.56 \pm 0.60\, (\pm 3.85)$ & $102.30 \pm 1.25\, (\pm 2.50)$ & \nodata & \nodata\\
 \cutinhead{MM5B-(a+c+d); $(l,b) = (-5.0, -3.5)$; $N_{\textrm{stars}} = 1193$}
 Likelihood  & $-56.46 \pm 1.84\, (\pm 4.48)$ & $102.04 \pm 0.09\, (\pm 0.28)$ & $-0.018 \pm 0.007\, (\pm 0.018)$
 & $-0.031 \pm 0.005\, (\pm 0.012)$ \\
 Biweight  & $-58.57 \pm 0.48 \, (\pm 3.35)$ & $101.46 \pm 1.21\, (\pm 2.03)$ & \nodata & \nodata \\
 \cutinhead{MM7B-(a+c+d); $(l,b) = (5.5, -3.5)$; $N_{\textrm{stars}} = 738$}
 Likelihood & $35.73 \pm 7.20\,(\pm 9.44)$ & $102.40 \pm 0.15\,(\pm 0.31)$ & $-0.016 \pm 0.031\,(\pm 0.043)$
 & $-0.047 \pm 0.022\, (\pm 0.026)$\\
 Biweight & $30.73 \pm 0.58\, (\pm 3.89)$ & $102.38 \pm 1.21\, (\pm 2.39)$ & \nodata & \nodata \\
 \cutinhead{BWC-a; $(l,b) = (1.06, -3.9)$; $N_{\textrm{stars}} = 557$}
 Likelihood  & $4.32 \pm 0.26\,(\pm 5.21)$ & $110.29 \pm 0.36\,(\pm 0.73)$ & $-0.031 \pm 0.011\, (\pm 0.019)$
 & $-0.035 \pm 0.009\, (\pm 0.017)$\\
 BW (B)  & $-1.01 \pm 5.42$ & $112.62 \pm 3.41$ & \nodata & \nodata \\
 NGC 6522 (N) & $-14.67 \pm 3.77$ & $7.9 \pm 2.82$ & & \\
 \enddata
 \tablecomments{There are two types of errors for the velocity moments: Gaussian
  randomization and bootstrap (in parenthesis).\\
 For BW kinematics the B and N stand for the broad and narrow Gaussian components.}
\end{deluxetable}

\begin{deluxetable}{ccrcccr}
 \tablecaption{Radial streaming motions.\label{streamdata}}
 \tablehead{\colhead{LOS} & \colhead{I$_{\mathrm{peak}}$} & \colhead{$\langle V_{\textrm{bright}}\rangle$} & \colhead{$N_{\textrm{bright}}$}
 & \colhead{$\langle V_{\textrm{faint}}\rangle$} & \colhead{$N_{\textrm{faint}}$} & \colhead{$\Delta V$}\\
 \colhead{(l,b)} & \colhead{} & \colhead{(\kms)} & \colhead{} & \colhead{(\kms)} & \colhead{}  & \colhead{(\kms)}}
 \startdata
 $(+5.5, -3.5)$ & 15.4 & $13.57 \pm 7.5$ & 186 & $52.90 \pm 8.3$ & 140 &  $-39.69 \pm 11.2 $  \\
 $(+1.06, -3.9)$ & 15.2 &$-7.05 \pm 11.3$ & 94 &  $-2.64 \pm 10.7$ & 109 & $-4.41 \pm 15.5$ \\
 $(-5.0, -3.5)$ & 15.6 & $ -73.48 \pm 8.1$ & 174 & $41.81 \pm 8.1$ & 167 & $-31.67 \pm 11.5$ \\
 \enddata
\end{deluxetable}

\end{document}